\documentclass[iop]{emulateapj}
\usepackage{amssymb}

\newcommand{\WMAP}{\textsl{WMAP}}

\newcommand{\iWMAP}{ {\it WMAP} }
\newcommand{\be}{\begin{equation}}
\newcommand{\ee}{\end{equation}}
\newcommand{\ba}{\begin{eqnarray}}
\newcommand{\ea}{\end{eqnarray}}
\newcommand{\nn}{\nonumber}
\newcommand{\barr}{\begin{array}}
\newcommand{\earr}{\end{array}}

\def\threej#1#2#3#4#5#6{\left( \begin{array}{ccc} #1 & #2 & #3 \\ #4 & #5 & #6 \end{array} \right) }
\def\n{{\bf \hat n}}
\def\L{{\mathcal L}}
\slugcomment{Accepted by the Astrophysical Journal Supplement Series}

\begin{document}

\title{Seven-Year {\it Wilkinson Microwave Anisotropy Probe} (\iWMAP\altaffilmark{1}) 
Observations: \\ 
Are There Cosmic Microwave Background Anomalies?}

\author{
{{C. L. Bennett}}\altaffilmark{2}, 
{{R. S. Hill}}\altaffilmark{3}, 
{{G. Hinshaw}}\altaffilmark{4},
{{D. Larson}}\altaffilmark{2}, 
{{K. M. Smith}}\altaffilmark{5}, 
{{J. Dunkley}}\altaffilmark{6}, 
{{B. Gold}}\altaffilmark{2},  
{{M. Halpern}}\altaffilmark{7}, 
{{N. Jarosik}}\altaffilmark{8},  
{{A. Kogut}}\altaffilmark{4}, 
{{E. Komatsu}}\altaffilmark{9}, 
{{M. Limon}}\altaffilmark{10}, 
{{S. S. Meyer}}\altaffilmark{11}, 
{{M. R. Nolta}}\altaffilmark{12}, 
{{N. Odegard}}\altaffilmark{3}, 
{{L. Page}}\altaffilmark{8}, 
{{D. N. Spergel}}\altaffilmark{5,13},  
{{G. S. Tucker}}\altaffilmark{14}, 
{{J. L. Weiland}}\altaffilmark{3}, 
{{E. Wollack}}\altaffilmark{4}, 
{{E. L. Wright}}\altaffilmark{15}}

\altaffiltext{1}{\WMAP\ is the result of a partnership between Princeton
University and NASA's Goddard Space Flight Center. Scientific
guidance is provided by the \iWMAP Science Team.}
\altaffiltext{2}{{Department of Physics \& Astronomy, %
                    The Johns Hopkins University, 3400 N. Charles St., %
                    Baltimore, MD  21218-2686, USA}}
\altaffiltext{3}{{ADNET Systems, Inc., %
                    7515 Mission Dr., Suite A100 Lanham, Maryland
                    20706, USA}}
\altaffiltext{4}{{Code 665, NASA/Goddard Space Flight Center, %
                    Greenbelt, MD 20771, USA}}
\altaffiltext{5}{{Department of Astrophysical Sciences, %
                    Peyton Hall, Princeton University, Princeton, NJ
                    08544-1001, USA}}
\altaffiltext{6}{{Astrophysics, University of Oxford, %
                    Keble Road, Oxford, OX1 3RH, UK}}
\altaffiltext{7}{{Department of Physics and Astronomy, University of %
                    British Columbia, Vancouver, BC  Canada V6T 1Z1}}
\altaffiltext{8}{{Department of Physics, Jadwin Hall, %
                    Princeton University, Princeton, NJ 08544-0708,
                    USA}}
\altaffiltext{9}{{University of Texas, Austin, Department of Astronomy, %
                    2511 Speedway, RLM 15.306, Austin, TX 78712, USA}}
\altaffiltext{10}{{Columbia Astrophysics Laboratory, %
                    550 W. 120th St., Mail Code 5247, New York, NY
                    10027-6902, USA}}
\altaffiltext{11}{{Depts. of Astrophysics and Physics, KICP and EFI, %
                    University of Chicago, Chicago, IL 60637, USA}}
\altaffiltext{12}{{Canadian Institute for Theoretical Astrophysics, %
                    60 St. George St, University of Toronto, %
                    Toronto, ON  Canada M5S 3H8}}
\altaffiltext{13}{{Princeton Center for Theoretical Physics, %
                    Princeton University, Princeton, NJ 08544, USA}}
\altaffiltext{14}{{Department of Physics, Brown University, %
                    182 Hope St., Providence, RI 02912-1843, USA}}
\altaffiltext{15}{{UCLA Physics \& Astronomy, PO Box 951547, %
                    Los Angeles, CA 90095-1547, USA}}

\begin{abstract}
A simple six-parameter $\Lambda$CDM model provides a successful fit to  
{\it WMAP} data.  This holds both when the \iWMAP data are analyzed alone or 
in combination with other cosmological data.  
Even so, it is appropriate to examine the data carefully to search for hints 
of deviations from the now standard model of cosmology, which includes inflation, 
dark energy, dark matter, baryons, and neutrinos.  
The cosmological community has 
subjected the {\it WMAP} data to extensive and varied analyses.  
While there is widespread agreement as to the overall success of the six-parameter 
$\Lambda$CDM model, 
various ``anomalies'' have been reported relative to that model.  In this paper we examine 
potential anomalies and present analyses and assessments of their significance.
In most cases we find that claimed anomalies depend on posterior selection of some aspect 
or subset of the data. Compared with sky simulations based on the best-fit model, 
one can select for low probability features of the {\it WMAP} data.  Low probability features 
are expected, but it is not usually straightforward to determine whether any particular 
low probability feature is the result of the {\it a posteriori} selection or non-standard 
cosmology.  Hypothesis testing
could, of course, always reveal an alternative model that is statistically favored, but there is 
currently no model that is more compelling.  
We find that two cold spots in the map are statistically consistent
with random cosmic microwave background (CMB)
fluctuations.   We also find
that the amplitude of the quadrupole is well within the expected 95\% confidence range 
and therefore is not anomalously low.  We find no significant anomaly with a lack of large 
angular scale CMB power for the best-fit $\Lambda$CDM model.  We examine in detail the 
properties of the power spectrum data with respect to the $\Lambda$CDM model and find no
significant anomalies.  The quadrupole and octupole components of the CMB sky are remarkably
aligned, but we find that this is not due to any single map feature; it results from the 
statistical combination of the full-sky anisotropy fluctuations.  It may be due, in part, to 
chance alignments between the primary and secondary anisotropy, but this only shifts the 
coincidence from within the last scattering surface to between it and the local matter 
density distribution.  
While this alignment appears to be remarkable, there was no model that 
predicted it, nor has there been a model that provides a compelling retrodiction. 
We examine claims of a hemispherical or dipole power 
asymmetry across the sky and find that the evidence for these claims is not statistically 
significant.  We confirm 
the claim of a strong quadrupolar power asymmetry effect, but there is considerable evidence 
that the effect is not cosmological.  The likely explanation is an insufficient handling 
of beam asymmetries.
We conclude that there is no compelling evidence for deviations from the  
$\Lambda$CDM model, which is generally an acceptable statistical fit to {\it WMAP} and 
other cosmological data.  
\end{abstract}


\section{Introduction}

The {\it WMAP} mission \citep{bennett/etal:2003a}
was designed to make precision measurements of the cosmic microwave
background (CMB) to 
place constraints on cosmology.  \iWMAP\ was specifically designed to minimize systematic
measurement errors so that the resulting measurements would be highly reliable within 
well-determined and well-specified uncertainty levels.  
The rapidly switched and highly symmetric differential radiometer system 
effectively makes use of the sky as a stable reference load and renders most potential systematic 
sources of error negligible.  The spacecraft spin and precession paths on the sky create a 
highly interconnected set of differential data.  Multiple radiometers and multiple frequency 
bands enable checks for systematic effects associated with particular radiometers and  
frequency dependencies.  Multiple years of observations allow for checks of time-dependent 
systematic errors.

The \iWMAP\ team has provided the raw time ordered data to the community. It has also made 
full-sky maps from these data, and these maps are the fundamental data product of the mission.  
If (and only if) the full-sky CMB anisotropy represented in a map is a realization of an isotropic 
Gaussian random process, then the power spectrum of that map contains all of the cosmological 
information.  The maps and cosmological parameter likelihood function based on the 
power spectrum are the products most used by  
the scientific community.  

The \iWMAP\ team used realistic simulated time ordered data to test
and verify the map-making process.  The \iWMAP\ and {\it Cosmic
Background Explorer}) ({\it COBE}) 
maps, produced by independent hardware
and with substantially different orbits and sky scanning patterns  have been found to be statistically 
consistent.  \cite{freeman/etal:2006} directly verified the fidelity of the \iWMAP\ team's 
map-making process \citep{hinshaw/etal:2003, jarosik/etal:2007, jarosik/etal:prep}.
It was indirectly verified by \cite{wehus/etal:2009} as well.  Finally, numerous CMB experiments 
have verified the \iWMAP\ sky maps over small patches of the sky (mainly with cross-correlation
analyses), either to extract signal or to transfer the more precise \iWMAP\ calibration. 

\iWMAP data used alone are consistent with a six-parameter inflationary $\Lambda$CDM model that specifies
the baryon density $\Omega_b h^2$, the cold dark matter density $\Omega_c h^2$, a cosmological 
constant $\Omega_\Lambda$, a spectral index of scalar fluctuations $n_s$, 
the optical depth to reionization $\tau$, and the scalar fluctuation amplitude $\Delta_{\cal R}^2$ 
\citep{dunkley/etal:2009, larson/etal:prep}.  This $\Lambda$CDM model is flat, with a nearly 
(but not exactly) scale-invariant fluctuation spectrum seeded by inflation, 
with Gaussian random phases, 
and with statistical isotropy over the super-horizon sky. 
When \iWMAP data are combined with additional cosmological data, the $\Lambda$CDM model remains 
a good fit,  
with a narrower range of allowed parameter values \citep{komatsu/etal:prep}. It is remarkable that 
such diverse observations over a wide range of redshifts 
are consistent with the standard $\Lambda$CDM model.

There are three major areas of future investigation: (1) further constrain allowed 
parameter ranges; (2) test the standard $\Lambda$CDM model against data to seek reliable 
evidence for flaws; and (3) seek the precise physical nature of 
the components of the $\Lambda$CDM model: cold dark matter, inflation, and dark energy.
It is the second item that we address here: are there potential deviations from 
$\Lambda$CDM within the context of the allowed parameter ranges of the existing \iWMAP\ 
observations?  

A full-sky map $T(\mathbf{n})$ may be decomposed into spherical harmonics
$Y_{lm}$ as
\begin{equation}
T(\mathbf{n})=\sum_{l=0}^\infty  \sum_{m=-l}^l  a_{lm} Y_{lm} (\mathbf{n})
\end{equation} 
with 
\begin{equation}
a_{lm} = \int d\mathbf{n} \: T(\mathbf{n}) Y_{lm}^* (\mathbf{n})
\end{equation} 
where $\mathbf{n}$ is a unit direction vector.  If the CMB anisotropy is Gaussian-distributed 
with random phases, then each $a_{lm}$ is independent, with a zero-mean 
$\left< a_{lm} \right>=0$ Gaussian distribution with
\begin{equation}
\left< a_{lm} a_{l^\prime m^\prime}^* \right> = \delta_{l l^\prime}  \delta_{m m^\prime} C_l
\end{equation} 
where $C_l$ is the angular power spectrum and $\delta$ is the Kronecker delta.  $C_l$ is the 
mean variance per multipole moment $l$ that would be obtained if one could take and average  
measurements from every vantage point throughout the universe.  We have only our one sample 
of the universe, however, and its spectrum is related to the measured $a_{lm}$ coefficients by
\begin{equation}
C_l^{\rm sky} = \frac{1}{2l+1} \sum_{m=-l}^l \vert a_{lm}\vert^2
\end{equation} 
where $\left< C_l^{\rm sky}\right>=C_l$ if we were able to average over an ensemble of
vantage points.  There is an intrinsic cosmic variance of 
\begin{equation}
\frac{\sigma_l}{C_l} = \sqrt{\frac{2}{2l+1}}. 
\end{equation}
In practice, instrument noise and sky masking complicate these relations.
In considering potential deviations from the $\Lambda$CDM model in this paper, we examine 
the goodness-of-fit of the $C_l$ model to the data, the Gaussianity of the $a_{lm}$ 
derived from the map, and correlations between the $a_{lm}$ values.  

We recognize that some versions of $\Lambda$CDM (such as with multi-field inflation, 
for example) predict a weak deviation from Gaussianity.  To date, the \iWMAP\ team has 
found no such deviations from Gaussianity.  This topic is further 
examined  by \cite{komatsu/etal:prep,
komatsu/etal:2009}.  Statistical isotropy is a key prediction of the simplest inflation 
theories so any evidence of 
a violation of rotational invariance would be a significant challenge to the 
$\Lambda$CDM or any model based on standard inflation models. 

Anomaly claims should be tested for contamination by systematic errors and foreground emission,
and should be robust to statistical methodology.
Statistical analyses of \iWMAP CMB data can be complicated, and 
simulations of skies with known properties are usually a necessary part of the analysis.
Statistical analyses must account for {\it a posteriori} bias, which is easier said than done.  
With the large amount of \iWMAP\ data and an enormous number of possible ways to combine the 
data, some number of low probability outcomes are expected.  For this reason, what 
constitutes a ``significant'' deviation from $\Lambda$CDM can be difficult to specify.  
While methods to reduce foreground
contamination (such as sky cuts, internal linear combinations, and template-based subtractions) 
can be powerful, none is perfect.  Since claimed anomalies often tend to be at marginal levels of
significance (e.g., $2-3\sigma$), the residual foreground level may be a significant consideration.  

The \iWMAP Science Team has searched for a 
number of different potential systematic effects and placed quantitative
upper limits on them.  
The \iWMAP team has extensively examined systematic measurement errors with each of its data 
releases:  \cite{jarosik/etal:2003} and \cite{hinshaw/etal:2003} for the first-year data release, 
\cite{hinshaw/etal:2007} and \cite{page/etal:2007} for the three-year data release, 
\cite{hinshaw/etal:2009} for the five-year data release, and \cite{jarosik/etal:prep} 
for the current seven-year data release.  Since those papers already convey the 
extensive systematic error analysis efforts of the \iWMAP team, this paper focuses on 
the consistency of the data with the $\Lambda$CDM model 
and relies on the systematic error limits placed in those papers.
Some data analysis techniques compute complicated combinations of 
the data where the systematic error limits must be fully propagated.  

This is one of a suite of papers presenting the seven-year {\it WMAP} data.  
\cite{jarosik/etal:prep}
provide a discussion of the sky maps, systematic errors, and basic results.  
\cite{larson/etal:prep}
derive the power spectra and cosmological parameters from the \iWMAP data.
\cite{gold/etal:prep}
evaluate the foreground emission and place limits on the foreground contamination 
remaining in the separated CMB data.
\cite{komatsu/etal:prep} present a cosmological interpretation of the \iWMAP data 
combined with other cosmological data.
\cite{weiland/etal:prep}
analyze the \iWMAP observations of the outer planets and selected bright sources, 
which are useful both
for an understanding of the planets and for enabling these objects to serve as more 
effective calibration sources for CMB and other millimeter-wave and microwave 
experiments.

This paper is organized as follows.
In Section \ref{coldspot1} we comment on the prominent large cold spot, nearby but offset 
from the Galactic Center region, that attracted  attention when the first \iWMAP\ sky map 
was released in 2003.  
In Section \ref{coldspot2} we comment on a cold spot in the southern sky that has attracted 
attention more recently.
In Section \ref{quad} we assess the level of significance of the low value measured for the 
amplitude of the CMB quadrupole ($l=2$) component.
In Section \ref{largescalepower} we discuss the lack of large-scale 
power across the sky.
In Section \ref{chi2} we assess the goodness-of-fit of the \iWMAP\ data to the $\Lambda$CDM model.
In Section \ref{alignedquad} we examine the alignment of the quadrupole and octupole.
In Section \ref{hemispherical} we assess claims of a hemispherical or dipole power asymmetry,
and in Section \ref{quadrupolar} we assess claims of a quadrupolar power asymmetry.
We summarize our conclusions in Section \ref{conclusions}.

\section{Cold Spot I, Galactic Foreground Emission, and the Four
Fingers}\label{coldspot1}

\begin{figure}[htp]
\epsscale{0.8}
\plotone{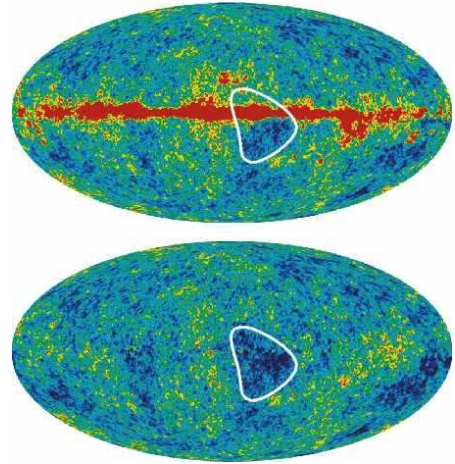}
\caption{Top: A large colder-than-average region, highlighted by the white curve, appears 
prominently on the raw V-band temperature map.  The full-sky map is shown with the Galactic plane 
horizontal across the center of the map with the Galactic center at the center of the 
displayed projection.  Bottom: With the foreground signals strongly suppressed by the ILC
technique, the highlighted cold spot is seen to be at least as prominent.  It is offset from the
Galactic center in both latitude and longitude.  This fact, combined with the fact that the  
clearly effective ILC foreground reduction does not diminish this feature, establishes that 
this is a CMB fluctuation and not a foreground effect.  This feature is not anomalous in that 
simulated realizations of $\Lambda$CDM model skies routinely produce features like this. 
\label{cold_spot_1}}
\end{figure}

\begin{figure}[htp]
\epsscale{1.0}
\plotone{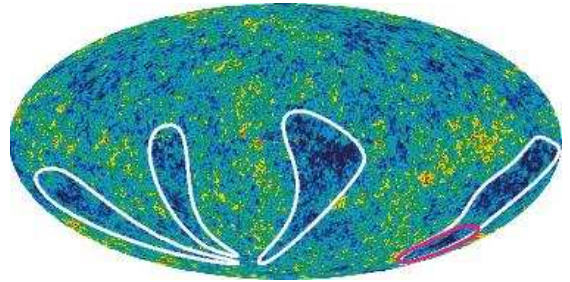}
\caption{  
Visual inspection of the ILC map reveals four elongated valleys of cooler temperature 
that stretch from about the Galactic equator to nearly the south Galactic pole.  Ridges of 
warmer-than-average temperature lie between the cooler fingers.  These
features are a consequence of 
large-scale power in the southern sky.  It is more difficult to discern as much large-scale 
power in the northern sky.  Cold Spot I is located near the northernmost part of one of the 
fingers, while Cold Spot II (within the red curve) is near the southernmost part of another 
finger. 
\label{fingers}}
\end{figure}

When \iWMAP data were first released in 2003, an image of the full sky was presented
in Galactic coordinates, centered on the Galactic center \citep{bennett/etal:2003b}.  
Galactic emission was 
minimized for this image by using an Internal Linear Combination (ILC) of  
\iWMAP data from independent frequency bands in such a way as to minimize signals with the 
frequency spectra of the Galactic foregrounds.  The seven-year raw V-band
map and ILC map are shown in Figure \ref{cold_spot_1}.  
A prominent cold (blue) spot is seen near the center of these maps, roughly half of 
which lies within the KQ85y7 Galaxy mask.  For the portion within the mask, the ILC 
process removes $>99$\% of the V-band pixel--pixel variance, while making almost no 
difference to the variance in the portion outside the mask. In the ILC map, the 
variances in the two regions are nearly equal.  Given its CMB-like spectrum and the 
fact that it is not centered on the Galactic center, this cold spot is very unlikely 
to be due to galactic foreground emission.
Several years of modeling the separation of foreground emission from the CMB 
emission continues to support the conclusion that this cold spot is not dominated 
by Galactic foreground emission, but rather is a fluctuation of the 
CMB.  
Further, the probability that a randomly located feature of this size would be 
near the Galactic 
center is $\sim 5$\% and the probability of such a feature overlapping the Galactic 
plane, at any longitude, is much higher.
This large central cold spot is a statistically reasonable CMB fluctuation within the context 
of the $\Lambda$CDM model.

Since people are highly effective at detecting patterns, it is not surprising that a visual 
inspection of the {\it WMAP} sky map reveals interesting features.  Four elongated 
cold (blue) fingers stretching from about the 
Galactic equator to the south Galactic pole are seen in Figure \ref{fingers}.  
There do not appear to be any similar fingers or features in the northern Galactic 
hemisphere aside from the northern-most extensions of the mostly southern fingers.  
Cold Spot I can be seen to be the northern part of one of the colder fingers.

There may be a tendency to overestimate the significance of features like Cold Spot I 
or the four fingers. 
It is very hard to define quantitative statistics for such features due to the 
visual nature of their identification. There is also an unavoidable posterior bias 
when using narrowly defined statistics targeted at particular features seen in our 
sky. In any case, visual inspection of simulated $\Lambda$CDM maps
often reveals large-scale features such as these, 
without requiring any underlying statistical fluctuation. 
Indeed, it is the lack of these features in the Northern sky that may be the more 
unusual situation.

\section{Cold Spot II}\label{coldspot2}

A detection of non-Gaussianity and/or phase correlations in the \iWMAP $a_{lm}$ data 
would be of great interest.  
While a detection of non-Gaussianity could be indicative of an experimental systematic effect 
or of residual foregrounds, it could also point to new cosmological physics. 
There is no single preferred test for non-Gaussianity.  
Rather, different tests probe different types of non-Gaussianity; therefore, it is important 
to subject the data to a variety of tests.

\cite{vielva/etal:2004} used a spherical Mexican hat wavelet (SMHW) analysis on the first year 
\iWMAP data to claim a detection of a non-Gaussian signal on a scale of a few degrees, 
independent of the \iWMAP observing frequency.
Also applying the SMHW kernel, 
\cite{mukherjee/wang:2004} claimed to detect non-Gaussianity at $\sim 99$\% significance.
The signal is a positive kurtosis in the wavelet coefficients attributed to a larger than 
expected number of 3\arcdeg\ to 5\arcdeg\ cold spots in the southern Galactic hemisphere.
\cite{mukherjee/wang:2004} found the same result for the ILC map.
Following up on this, \cite{cruz/etal:2006} reported that the kurtosis in the wavelet 
distribution could be exclusively attributed to a single cold spot, which we call Cold Spot II, 
in the sky map at 
Galactic coordinates ($l = 209^\circ$, $b=-57^\circ$), as indicated by the red curve in 
Figure \ref{fingers}.   
In an analysis of the three-year \iWMAP data, \cite{cruz/etal:2007a} reported only 1.85\% of their 
simulations deviated from the \iWMAP data either in the skewness or in the kurtosis estimator 
at three different angular scales, a $2.35\sigma$ effect.  

In replicating the SMHW approach described above, \cite{zhang/huterer:2010} found $\sim 2.8\sigma$ 
evidence of kurtosis and $\sim 2.6\sigma$ evidence for a cold spot. These values are in reasonable
agreement with the earlier findings for individual statistics.  
Zhang and Huterer then allowed for a  range of possibilities 
in disk radius, spatial filter shape, and the choice of non-Gaussian statistic in a 
``superstatistic''.  They found that 23\% of simulated Gaussian random skies are more unusual than 
the \iWMAP sky.  \cite{zhang/huterer:2010} also analyzed the sky maps with circular disk and Gaussian 
filters of varying widths. They found no evidence for an anomalous cold spot at any scale when 
compared with random Gaussian simulations.  When requiring the SMHW spatial filter shape, 15\% of 
simulated Gaussian random skies were more unusual than the \iWMAP sky
using the constrained superstatistic.   

With 1.85\% -- 15\% of random Gaussian skies more deviant than \iWMAP data in a wavelet analysis  
(depending on the marginalization of posterior choices) potential physical interpretations have 
been proposed for this $1.45\sigma$ -- $2.35\sigma$ effect.
In theory, cold spots in the CMB can be produced by the integrated
Sachs--Wolfe (ISW) effect as 
CMB photons traverse cosmic voids along the line of sight. If Cold Spot II is due to a cosmic 
void, it would have profound implications because $\Lambda$CDM does not produce voids of 
sufficient magnitude to explain it. \cite{mota/shaw/silk:2008} examined void formation in models 
where dark energy was allowed to cluster and concluded that voids of sufficient size to explain 
Cold Spot II were not readily produced. \cite{rudnick/etal:2007} examined number counts and smoothed 
surface brightness in the NRAO VLA Sky Survey (NVSS) radio source
data. They claimed a 20\% -- 45\% 
deficit in the NVSS smoothed surface brightness in the direction of Cold Spot II. However, 
this claim was refuted by \cite{smith/huterer:2010}, who found no evidence for a deficit, 
after accounting for systematic effects and posterior choices made in assessing statistical 
significance. Further, \cite{granett/etal:2010} imaged several fields within Cold Spot II on the 
Canada--France--Hawaii Telescope and attained galaxy counts that rule out a 100 Mpc radius 
spherical void at high significance, finding no evidence for a supervoid. 

\cite{cruz/etal:2007b} suggested that the cold spot could be the signature of a topological defect 
in the form of a cosmic texture rather than an adiabatic fluctuation.  
This suggestion was further discussed by \cite{bridges/etal:2008} 
and \cite{cruz/etal:2008}: they estimate that a texture with $G\mu \approx 1.5\times 10^{-6}$ could 
produce a 
cold spot with the observed properties. Independently, CMB power spectra combined with other 
cosmological data can 
be used to place limits on a statistical population of topological defects.  
\cite{urrestilla/etal:2008} placed a 95\% 
confidence upper limit of $G\mu < 1.8\times 10^{-6}$ based on Hubble constant, nucleosynthesis, 
and CMB (including three-year \iWMAP) data.  Textures at this level are compatible with the
cold spot and are neither favored nor disfavored by parameter fits.  Following the method 
described in 
\cite{urrestilla/etal:2008}, we now place a power spectrum based 95\% confidence upper limit of 
$G\mu < 1.5\times 10^{-6}$ using the seven-year \iWMAP data, finer scale CMB data, and the 
Hubble constant.  
Since the new 95\% CL upper limit derived from the power spectrum matches the 
value needed to explain the cold spot within the simple texture model previously discussed 
in the literature, that model  is now disfavored; however, a more sophisticated model might 
still allow a texture interpretation.

In conclusion, there are two possible points of view.  One is that the cold spot is anomalous 
and might be produced by a texture or other mechanism; this cannot be ruled out with current data.
The other view is that the $<2.35\sigma$ (after
posterior correction) statistical evidence for a cold spot feature is not compelling, and that 
the texture explanation is disfavored.
Had the anomaly been significant at the part per million level instead 
of a part per thousand, the posterior marginalization issues would be moot: the feature would 
be considered a real anomaly.  
If real, a void explanation would have been in strong conflict with the $\Lambda$CDM model, but 
a population of topological defects that contribute at a low level to the CMB power
spectrum would not so much falsify the model as provide a small modification to it.  Given the 
evidence to date, the \iWMAP Team is of the opinion that there is insufficient statistical support 
to conclude that the cold spot is a CMB anomaly relative to $\Lambda$CDM.

\begin{figure}[htp]
\epsscale{1.0}
\plotone{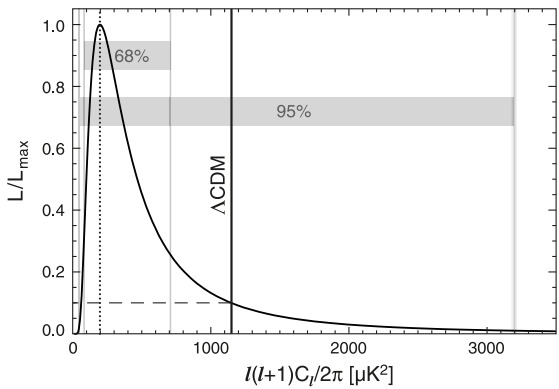}
\caption{
Curve is a Blackwell-Rao estimate of the relative likelihood of the quadrupole 
power $l(l+1)C_2/2\pi$ in $\mu$K$^2$ from \iWMAP.  The \iWMAP\ ILC data were smoothed to 
$5^\circ$ and the KQ85y7 mask was used, both degraded to res 5.  The Gibbs sampling produced 
a likelihood that has been marginalized over all other multipoles.    
The highly non-Gaussian likelihood distribution is characteristic of the lowest-$l$ multipoles.  
For $l > 32$ the curves become nearly Gaussian. 
The vertical line with the label ``$\Lambda$CDM'' is the expected quadrupole from the full power 
spectrum $\Lambda$CDM model best fit to the \iWMAP data.  The maximum likelihood of the 
\iWMAP\-measured $l=2$ quadrupole is at the vertical dotted line.  These two values are consistent 
to well within the 95\% confidence region.  The \iWMAP quadrupole is not anomalously low.  
\label{fig:quad_like}}
\end{figure}

\section{The Low Quadrupole Amplitude}\label{quad}

The CMB quadrupole is the largest observable structure in our universe. The magnitude of 
the quadrupole has been known to be lower than models predict since it was first measured by 
{\it COBE} \citep{bennett/etal:1992b, hinshaw/etal:1996a}. It is also the large-scale mode that is 
most prone to 
foreground contamination, owing to the disk-like structure of the Milky Way; thus estimates 
of the quadrupole require especially careful separation of foreground and CMB emission. The 
ILC method of foreground suppression is especially appropriate 
for large-scale foreground removal, since the ILC's complex small-scale noise properties are 
unimportant in this context. With a sky cut applied, residual foreground contamination of the quadrupole in 
the ILC map is determined to be insignificant \citep{gold/etal:prep, jarosik/etal:prep}. 
In this section we reassess the statistical significance of the low quadrupole power in 
the ILC map. 

A statistical analysis of the quadrupole must account for the highly non-Gaussian 
posterior distribution of the low-l ($l\lesssim 32$) multipoles.  
In this paper we use Gibbs sampling of the low-$l$ power spectrum to evaluate 
the quadrupole \citep{odwyer/etal:2004, dunkley/etal:2009}.  
The \iWMAP\ ILC data are smoothed to $5^\circ$ resolution (Gaussian, FWHM), degraded 
to HEALPix resolution level 5 ($N_{\rm side}=32$), and masked with the KQ85y7 mask. 
The Gibbs sampling produces a likelihood that has been marginalized over all other 
multipoles.   
A Blackwell-Rao estimator of the form of Equation 19 of \cite{wandelt/etal:2004} 
is used to produce the \iWMAP quadrupole likelihood shown in Figure \ref{fig:quad_like}.
The peak of the likelihood is at 200 $\mu$K$^2$, the median is at 430 $\mu$K$^2$, and the 
mean is at 1000 $\mu$K$^2$.
The 68\% confidence range extends from 80 to 700 $\mu$K$^2$, and the 95\% confidence range 
extends from 40 to 3200 $\mu$K$^2$.  

\begin{figure}[htp]
\epsscale{1.0}
\plotone{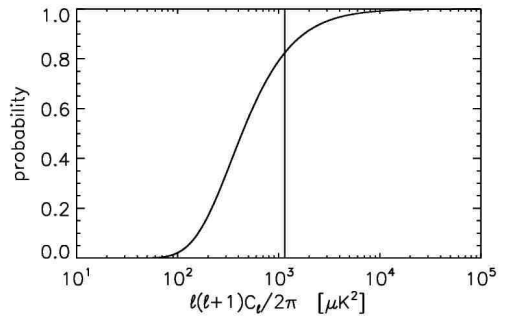}
\caption{
Cumulative distribution function of the quadrupoles from the Gibbs sampling based
on 300,000 points.
The vertical line is the predicted $\Lambda$CDM model quadrupole value.  
The cumulative probability is 0.824 where the vertical line crosses the 
cumulative distribution function . 
Since the expected quadrupole from the $\Lambda$CDM model is well within the 95\% confidence range 
of the measured quadrupole, accounting for noise and cosmic variance, we conclude that the 
measured quadrupole is not anomalously low. 
\label{fig:quad_cum_prob}}
\end{figure}

Figure \ref{fig:quad_cum_prob} shows the cumulative quadrupole distribution derived from 
300,000 Gibbs samples.  The mean quadrupole predicted by the best-fit $\Lambda$CDM model 
lies at a cumulative probability of 0.824, well within the 95\% confidence region allowed 
by the data. We conclude that the \iWMAP quadrupole measurement is not anomalously low. 
Further, while alternative models that predict a lower quadrupole will better match this 
specific part of the data, it is impossible to significantly favor such models on the basis of 
quadrupole power alone. 

\section{The Lack of Large-Scale Power}\label{largescalepower}

The angular correlation function complements the power spectrum by measuring structure 
in real space rather than Fourier space.  It measures the covariance of pixel temperatures 
separated by a fixed angle,
\begin{equation}
C(\theta_{ij})= \left< T_i T_j\right> 
\label{cpix}
\end{equation}
where $i$ and $j$ are two pixels on the sky separated by an angle $\theta$, and the 
brackets indicate an ensemble average over independent sky samples.  Expanding the 
temperature in spherical harmonics, and using the addition theorem for spherical harmonics, 
it is straightforward to show that $C(\theta_{ij})$ is related to the 
angular power spectrum by
\begin{equation}
\left< C(\theta_{ij})\right> = \frac{1}{4\pi} \sum_l (2l+1)\left< C_l\right> W_l P_l (\cos
\theta ) 
\label{cthetacl}
\end{equation}
where $\left< C_l\right>$ is the ensemble-average angular power spectrum, 
$W_l$ is the experimental 
window function, and $P_l$ are the Legendre polynomials.  If the CMB is statistically 
isotropic, $C(\theta_{ij}) \equiv C(\theta)$ depends only on the separation of pixels $i$ 
and $j$, but not on their individual directions.  Since we are unable to observe an ensemble 
of skies, we must devise estimates of $C(\theta)$ using the temperature measured in our sky.  
One approach is to assume that the CMB is ergodic (statistically isotropic) in which case 
$C(\theta)$ can be estimated by averaging all temperature pairs in the sky separated by an 
angle $\theta$,
\begin{equation}
C(\theta)= \left< T_i T_j\right> \vert_{\angle ij=\theta}
\label{cthetapix}
\end{equation}
where the brackets indicate an average over directions $i$ and $j$ such that $\angle ij=\theta$ 
(to within a bin).  Another approach is to estimate the angular power spectrum $C_l$ and to 
compute $C(\theta)$ using Equation (\ref{cthetacl}).

\begin{figure}[htp]
\epsscale{1.0}
\plotone{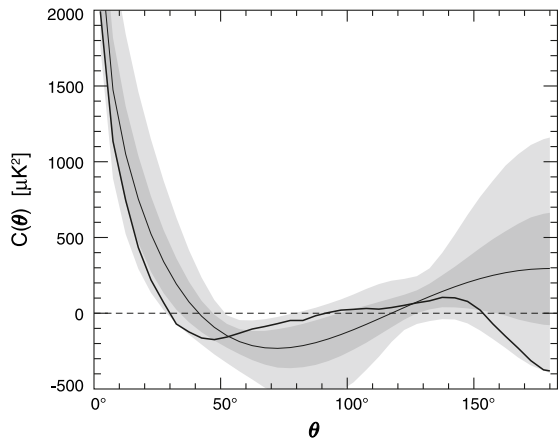}
\caption{Angular correlation function of the full-sky \iWMAP ILC map is shown (heavy 
black curve). For comparison, the angular correlation function for the best-fit 
$\Lambda$CDM model is also shown (thin black curve), along with the associated 68\% and 95\% 
confidence ranges, as determined by Monte Carlo simulations.  The angular correlation function 
of the full-sky map is seen to be within 
the 95\% confidence range of the best-fit $\Lambda$CDM model.  
This angular correlation function was computed from the $C_l$ power spectrum, but 
is nearly indistinguishable from a pixel pair computation. Either way, there is no evidence of a 
lack of large-scale power.
\label{coftheta}}
\end{figure}

The angular correlation function over the full-sky ILC map from Equation (\ref{cthetacl}) 
is shown in Figure \ref{coftheta}.  
As can be seen, $C(\theta)$ lies within the 95\% confidence range of the best-fit 
$\Lambda$CDM model for all $\theta$, as 
determined by Monte Carlo simulations.  This supports the conclusion that there is no 
statistically significant lack of large-scale power on the full sky.

\cite{spergel/etal:2003} applied the pixel-pair estimator to the first-year \iWMAP data and 
found an almost complete lack of correlated structure at angles $>60\arcdeg$ for the sky, 
but that calculation was  
with a Galactic foreground cut.  A foreground cut was made because of the concern that 
additional power from within the Galactic cut may arise from foregrounds.  For regions 
outside the cut, it was appreciated that 
systematic errors and residual Galactic foregrounds are far more likely to add 
correlated power to the sky maps than to remove it.  
They quantified the lack of large-angular-scale power 
in terms of the statistic 
\begin{equation}
S_{1/2}\equiv \int_{-1}^{1/2} C^2(\theta) d \cos \theta
\end{equation}
and found that fewer than 0.15\% of simulations had lower values of 
$S_{1/2}$.  A low $S_{1/2}$ value persists in later {\it WMAP} sky maps.

\cite{copi/etal:2007} and \cite{copi/etal:2009} claimed that there is evidence 
that the \iWMAP temperature fluctuations 
violate statistical isotropy.  They directly computed the angular 
correlation function from pixel pairs, as in Equation (\ref{cthetapix}),
omitting from the sum pixel pairs where at least one pixel was within the foreground mask.  
The KQ85 foreground mask (at that time) removed 18\% of the pixels from the full sky (now 22\% 
for KQ85y7), 
while KQ75 
removed 29\%.  Copi et al. found $p$-values of $\approx 0.03\%$ for their computation of 
$S_{1/2}$, concluding that the data are quite improbable given the model.  
The exact $p$-value depended on the specific choice of CMB map and foreground mask.
\cite{cayon:2010} finds no frequency dependence to the effect.

\cite{efstathiou/ma/hanson:2010} find that the value of $S_{1/2}$ is sensitive to the method of
computation.  For example, \cite{efstathiou/ma/hanson:2010} computed the angular correlation 
function using the estimator
\begin{equation}
C(\theta)=\frac{1}{4\pi}\sum_{ll^\prime} (2l+1) M_{ll^\prime}^{-1} \widetilde{C}_{l^\prime} P_l(cos\theta)
\end{equation}
where
\begin{equation}
M_{ll^\prime}=\frac{1}{2l+1}\sum_{mm^\prime}\vert K_{lml^\prime m^\prime}\vert^2
\end{equation}
and $K_{lml^\prime m^\prime}$ is the coupling between modes ($lm$) and ($l^\prime m^\prime$)
induced by the sky cut, and $\widetilde{C}_{l^\prime}$ is the
pseudo-power spectrum obtained 
by transforming the sky map into spherical harmonics on the cut sky.  This estimator produced 
a significantly larger value for $S_{1/2}$ than the estimator in Equation (\ref{cpix}).  

\cite{efstathiou/ma/hanson:2010} also reconstructed the low-$l$ multipoles across the
foreground sky cut region in a manner that was numerically stable, without an assumption of
statistical isotropy.  Their method relied on the fact that the low multipole \iWMAP\ data are 
signal-dominated and that the cut size is modest.  They showed that the small reconstruction 
errors introduce no bias 
and they did not depend on assumptions of statistical isotropy or Gaussianity.  The
reconstruction error only introduced a small ``noise'' to the angular correlation function 
without changing its shape.  

The original use of a sky cut in calculating $S_{1/2}$ was motivated by concern for residual 
foregrounds in the ILC map.  We now recognize that this precaution was unnecessary as the 
ILC foreground residuals are relatively  small.  Values of $S_{1/2}$ are smaller on the cut sky 
than on the full sky, but since the full sky contains the superior sample of the universe and 
the cut sky estimates suffer from a loss of information, cut sky estimates must be considered 
sub-optimal.
It now appears that the \cite{spergel/etal:2003} and \cite{copi/etal:2007, copi/etal:2009} 
low $p$-values result from both the 
{\it a posteriori} definition of $S_{1/2}$ and a chance alignment of the Galactic plane with 
the CMB signal.    The alignment involves Cold Spot I and the northern tips of the other fingers,
and can also be seen in the maps that will be discussed in Section \ref{alignedquad}.

\cite{efstathiou/ma/hanson:2010} corrected the full-sky \iWMAP ILC map for the estimated 
ISW signal from redshift $z<0.3$ as estimated by  
\cite{francis/peacock:2010}.  The result was a substantial increase in the $S_{1/2}$.
Yet there is no large-scale cosmological significance to the orientation of the sky cut 
or the orientation of the local distribution of matter with respect to us; thus the result 
from Spergel et al. and Copi et al. must be influenced by a chance alignment of the ISW 
effect and a posterior statistical bias in the choice of statistic.

More generally, \cite{hajian/souradeep/cornish:2005} applied their bipolar power spectrum technique  
and found no evidence for a violation of statistical isotropy at 95\% CL for angular 
scales corresponding to multipole moments $l<60$.  

The low value of the $S_{1/2}$ integral over the large-angle correlation 
function on the cut-sky results from a posterior choice of the statistic.  Further, 
it is a sub-optimal statistic in that it is not computed over the full sky.  There is evidence 
for a chance 
alignment of the Galactic plane cut with the CMB signal, and there is also evidence of a chance 
alignment of the primary CMB fluctuations with secondary ISW features from the local 
density distribution.  The full-sky angular correlation function lies within the 95\% confidence 
range.  For all of these reasons, we conclude that the large-angle CMB correlation 
function is consistent with $\Lambda$CDM.

\section{The Goodness of Fit of the Standard Model}\label{chi2}

\begin{figure}[htp]
\epsscale{0.7}
\plotone{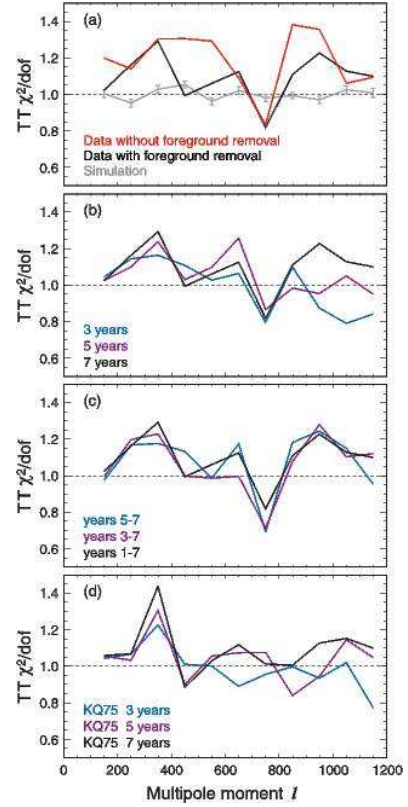}
\caption{
(a) $\chi^2$ per degree of freedom of the 
seven-year temperature-temperature power spectrum data 
relative to the best-fit $\Lambda$CDM model.
Light gray points are from 50 simulations that used the 
same $\Lambda$CDM model 
with the appropriate noise and cosmic variance included, where 
the error bars are driven by the number of simulations.  The 
data and simulations were run through the same data analysis 
pipeline.  These simulations were used to help fit the effective 
fraction of the sky, $f_{\rm sky}$ to use for the data analysis.
(b) The $\chi^2$ per degree of freedom is compared for the  
three-year, five-year, and seven-year maps.  Small differences in the fit model have a 
negligible effect on these plots.  The $\chi^2$ per degree of 
freedom for $l\approx 300$ has been slightly growing with additional data, 
while other multipole moment ranges are more random with additional data.
(c) The $\chi^2$ per degree of freedom had the same \iWMAP\ 
data been taken in a reverse time order.  The $l < 400$ region appears 
robust, while the $\chi^2$ variations for $l > 400$ appear more random.  
(d) Variation of $\chi^2$ with the choice of Galactic foreground mask 
appears random for $l>400$.  
\label{fig:chi2}}
\end{figure}

\begin{figure}[htp]
\epsscale{1.0}
\plotone{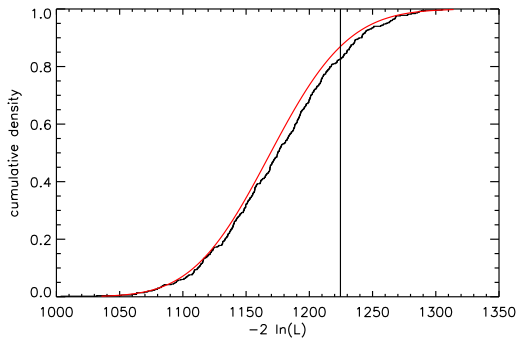}
\caption{Black curve is the cumulative probability of the \iWMAP\ 
temperature data based on 499 simulations.  All 
of the simulations were drawn from the same $\Lambda$CDM model, but $\chi^2$ was 
evaluated with respect to the best-fit model for each realization.
Of these, 412 (82.6\%) had a lower $\chi^2$ than the vertical 
line at 1224.6.
Thus, the \iWMAP\ power spectrum is statistically compatible with the 
model.  The red curve is a $\chi^2$ distribution with 1170 degrees of freedom, 
shown for comparison.
\label{chi2cum}}
\end{figure}

\begin{figure*}[!t]
\epsscale{0.8}
\plotone{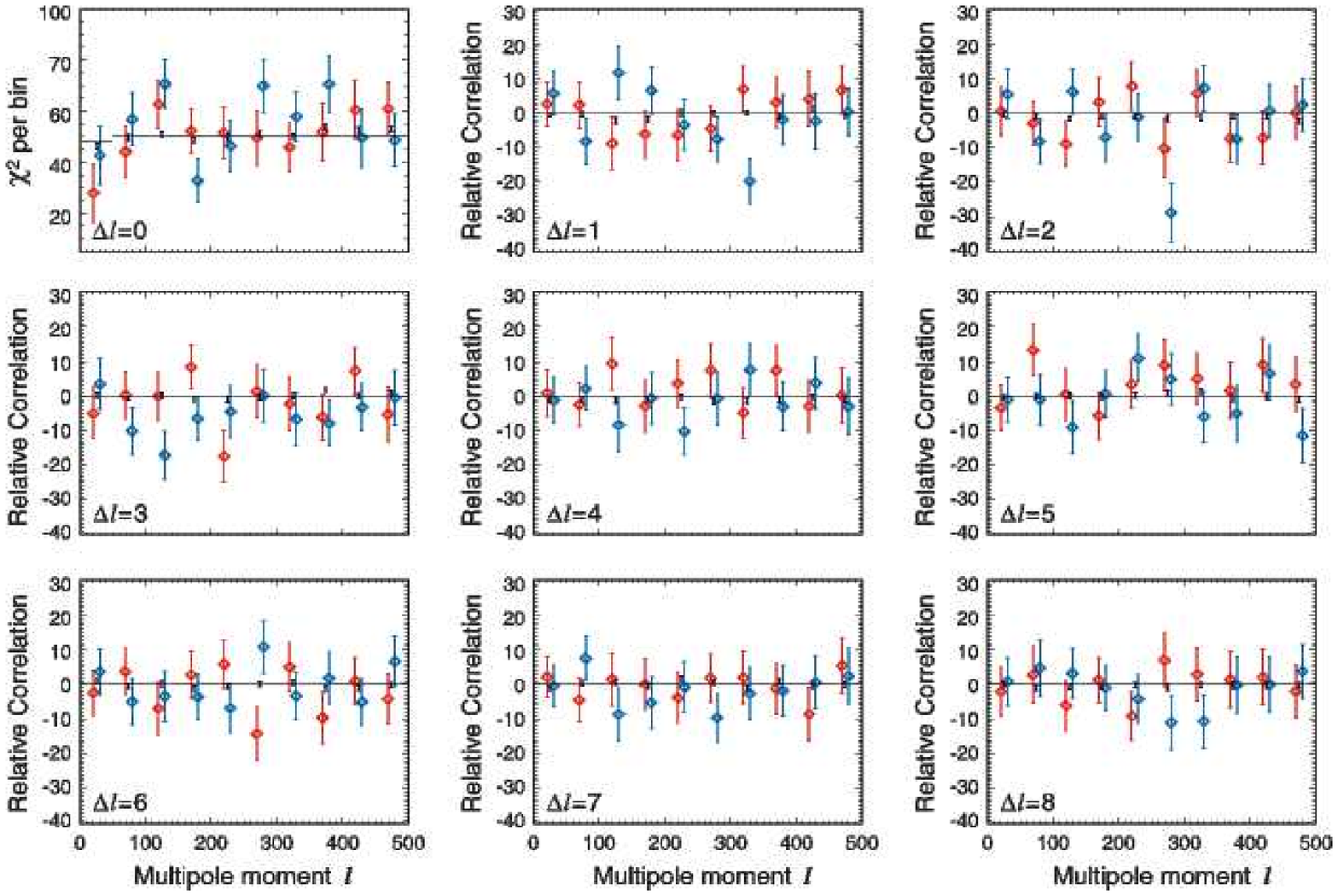} 
\caption{We compute $\Delta l=l-l'$ correlation properties of $C_l$ for nearby multipoles 
for the \iWMAP\ data (blue) and comparable simulations (red).  For the most part the data and simulations 
are in good agreement.  The most discrepant correlations in $C_l$ are for $\Delta l = 1$ 
near $l\sim 320$ and $\Delta l \sim 2$ near $l\sim 280$.
\label{clcorr}}
\end{figure*}

The power spectrum of the \iWMAP\ data alone places strong constraints on possible cosmological 
models \citep{dunkley/etal:2009, larson/etal:prep}.  
Plots of the $\chi^2$ per degree of freedom of the temperature--temperature power 
spectrum data relative to the best-fit $\Lambda$CDM model 
are shown in Figure \ref{fig:chi2}.  In Figure \ref{chi2cum} the cumulative 
probability of the \iWMAP\ data given the $\Lambda$CDM model is evaluated  
based on simulations.  All of the simulated skies were calculated for the same 
input $\Lambda$CDM model, but each result was fit separately. The \iWMAP\ sky is 
statistically compatible with the model within 82.6\% confidence, with an uncertainty of 
$\sim5$\%.

The $\chi^2$ can be elevated because of excess scatter within each multipole relative to the 
experimental noise variance.  
It could also be elevated because of an accumulation of systematic deviations of the 
model from the data across different multipoles, such as would happen
if a parameter value were
incorrect.   
Therefore, despite an acceptable overall $\chi^2$, we examine other aspects of the power 
spectrum data relative to the model that may have been masked by the inclusion of 
all of the data into a single $\chi^2$ value.  We examine both of these possibilities below.

To explore the $l$-to-$l'$ correlation properties of $C_l$, we compute:
\begin{eqnarray*}
 S(\Delta l) &\equiv&
\sum_{l}
\frac{(C_l^{\rm data}-C_l^{\rm bestfit})(C_{l+\Delta l}^{\rm data}-C_{l+\Delta l}^{\rm
bestfit})}{\sqrt{\frac{2(C_l^{\rm bestfit}+N_l)^2}{(2l+1)f^2_{{\rm
sky},l}}\frac{2(C_{l+\Delta l}^{\rm bestfit}+N_{l+\Delta l})^2}{(2l+2\Delta l+1)f^2_{{\rm
sky},l+\Delta l}}}}\\
&=&
\sum_l
f_{{\rm sky},l}f_{{\rm sky},l+\Delta l}
\sqrt{\left(l+\frac12\right)\left(l+\Delta l+\frac12\right)}\\
&&\times \frac{(C_l^{\rm data}-C_l^{\rm bestfit})(C_{l+\Delta l}^{\rm data}-C_{l+\Delta l}^{\rm
bestfit})}{(C_l^{\rm bestfit}+N_l)(C_{l+\Delta l}^{\rm bestfit}+N_{l+\Delta l})}.
\end{eqnarray*}
When $\Delta l=0$, this quantity is exactly $\chi^2$:  
$$
S(\Delta l=0)=\chi^2.
$$
For $l=300-349$ and $f_{\rm sky}\sim 0.8$, 
$$
S(\Delta l)\sim 200
\sum_{l=300}^{349}
\frac{(C_l^{\rm data}-C_l^{\rm bestfit})(C_{l+\Delta l}^{\rm data}-C_{l+\Delta l}^{\rm
bestfit})}{(C_l^{\rm bestfit}+N_l)(C_{l+\Delta l}^{\rm bestfit}+N_{l+\Delta l})}.
$$
Since the data suggest $S(\Delta l=1)\sim -20$ for $l=300-349$, we find
$$
\frac{(C_l^{\rm data}-C_l^{\rm bestfit})(C_{l+1}^{\rm data}-C_{l+1}^{\rm
bestfit})}{(C_l^{\rm bestfit}+N_l)(C_{l+1}^{\rm bestfit}+N_{l+1})}
\sim -0.002.
$$
For this multipole range $(C_l^{\rm bestfit}+N_l)\sim 3500~\mu{\rm
K}^2$; thus, 
$$
(C_l^{\rm data}-C_l^{\rm bestfit})(C_{l+1}^{\rm data}-C_{l+1}^{\rm
bestfit})
\sim -(160~\mu{\rm K}^2)^2.
$$
Note that the power spectrum of the \cite{finkbeiner/davis/schlegel:1999} dust map in 
this multipole range is $< 10~\mu{\rm K}^2$ in W band, i.e., more than an order of 
magnitude smaller.

Figure \ref{clcorr} shows the results of $l$-to-$l'$ correlation calculations of $C_l$ 
for different values of $l-l'$, calculated both for simulations and for the \iWMAP\ data.  
For the most part the data and simulations 
are in good agreement.  The most discrepant correlations in $C_l$ are for $\Delta l = 1$ 
near $l\sim 320$ and $\Delta l = 2$ near $l\sim 280$.

Motivated by the outlier $\Delta l=1$ correlation at $l\sim 320$ 
seen in Figure \ref{clcorr}, we further explore a possible even $l$
versus odd $l$ effect in this portion of the power spectrum.  
(Note that this is an {\it a posteriori} selection.)
We define an even excess statistic, $\mathcal{E_\ell}$, which compares 
the mean
power at even values of $\ell$ with the mean power at odd values
of $\ell$, within a given $\ell$-range. It is essentially a measure of
anticorrelation between adjacent elements of the power spectrum, with
a sign indicating the phase of the anticorrelation:
\begin{displaymath} 
\mathcal{E_\ell} = 
\frac{\langle \mathcal{C}^\mathrm{obs}_\ell 
- \mathcal{C}^\mathrm{th}_\ell\rangle_\mathrm{even}
- \langle \mathcal{C}^\mathrm{obs}_\ell 
- \mathcal{C}^\mathrm{th}_\ell\rangle_\mathrm{odd}}
{\langle \mathcal{C}^\mathrm{th}_\ell\rangle},
\end{displaymath}
where $\mathcal{C}_\ell = \ell(\ell+1)C_\ell/2\pi$, the superscript
``obs'' refers to the observed power spectrum, and the superscript
``th'' refers to a fiducial theoretical power spectrum used for
normalization.  
From this definition, it follows that $\mathcal{E_\ell}>0$ is an \emph{even excess} 
and $\mathcal{E_\ell}<0$ is an \emph{odd excess}.
The range of $\ell$ is small enough that the
variation in $\mathcal{C}^\mathrm{th}_\ell$ is also small and
convenient for normalization.  We choose ({\it a posteriori}) to bin
$\mathcal{E}_\ell$ by $\Delta\ell = 50$.  

An apparent positive $\mathcal{E}_\ell$ in
the \iWMAP\ power spectrum in the range $\ell\sim200-400$ is investigated
quantitatively using Monte Carlo simulations.  Our analysis is limited
to $ 33 \leq \ell < 600$, which is the part of the power spectrum that is 
flat-weighted on the sky and where the Monte Carlo Apodised Spherical 
Transform EstimatoR \citep{hivon/etal:2002} pseudo-spectrum is used.
Because the binning is by $\Delta\ell=50$, the actual $\ell$ range for
this analysis is $50-599$.  
The Monte Carlo realizations are CMB sky map simulations 
incorporating appropriate \iWMAP\ instrumental noise and beam smoothing.
Each power spectrum, whether from observed
data or from the Monte Carlo generator, is co-added from 861
year-by-year cross spectra in the V and W bands, with weighting that
accounts for the noise and the beam transfer functions.

\begin{figure}[htb]
  \epsscale{0.95}
  \plotone{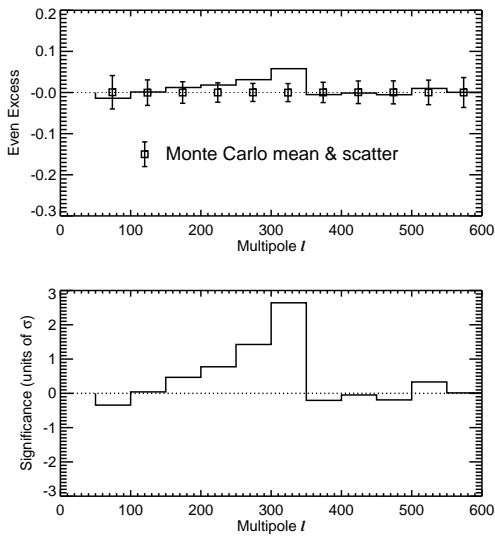}
  \caption{Top: even excess $\mathcal{E}_\ell$ in the observed power
  spectrum, in bins of $\Delta\ell=50$, compared to the mean and
  scatter from 11,436 Monte Carlo realizations.  Bottom:
  $\mathcal{E}_\ell$ as in the top plot, converted to significance
  units by normalizing to the Monte Carlo scatter in each bin.  Only
  the $\ell=250-299$ and $\ell=300-349$ bins show a significance
  greater than $1\sigma$.
    \label{fig:eebin}}
\end{figure}

\begin{figure}[htb]
  \epsscale{0.95}
  \plotone{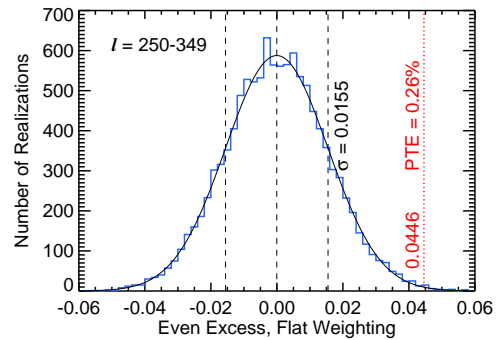}
  \caption{Even excess $\mathcal{E}_\ell$ in the observed power
  spectrum, in the $\ell$ range $250-349$, compared to a histogram
  of the same value as computed from 11,436 Monte Carlo realizations.
  The probability for a random realization to exceed the observed
  value (PTE) is $0.26\%$, interpolated directly from the Monte Carlo
  cumulative distribution.  A Gaussian with the same area and
  standard deviation as the histogram is overplotted, with $\pm 1\sigma$
  indications.
    \label{fig:eebigbin}}
\end{figure}

Figure \ref{fig:eebin} shows that an even excess of significance
$\sim2.7\sigma$ is found for $\ell=300-349$.  If we combine the two
adjacent bins between $\ell=250$ and $\ell=349$, the significance of
$\mathcal{E}_\ell$ in the combined bins is $\sim 2.9\sigma$, with a
probability to exceed (PTE) of $0.26\%$ integrated directly from the
Monte Carlo set (Figure \ref{fig:eebigbin}).  However, it is important 
to account for the fact that this
significance level is inflated by the posterior bias of having
chosen the $\ell$ range to give a particularly high value.

We attempt to minimize the posterior bias by removing bin selection
from the Monte Carlo test.  Instead of focusing on one bin, the
revised test is based on the distribution of the maximum value of
the significance $\mathcal{E}_\ell/\sigma(\mathcal{E}_\ell)$ 
over all bins in each Monte Carlo realization.  The 11436
Monte Carlo realizations are split into two groups:  4000 are used
to compute the normalization $\sigma(\mathcal{E}_\ell)$ for each bin,
and 7436 are used to compute the distribution of the maximum value
of $\mathcal{E}_\ell/\sigma(\mathcal{E}_\ell)$, giving the histogram
that is compared to the single observed value.

\begin{figure}[htp]
\epsscale{0.85}
  \plotone{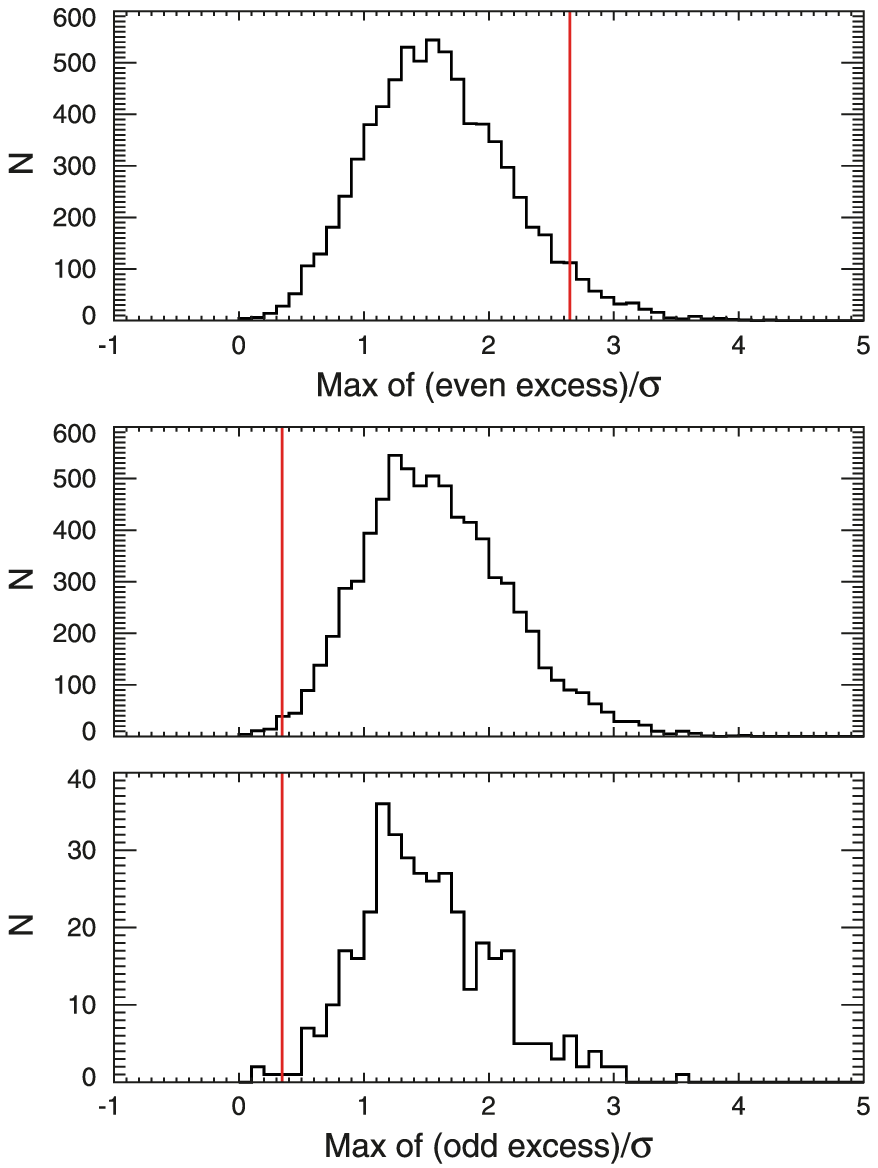}
  \caption{Top:  histogram of $\max\left(\mathcal{E}_\ell/\sigma(\mathcal{E}_\ell)\right)$
  for each Monte Carlo realization.  The maximum is taken over 11 bins
  of $\Delta\ell=50$, for $\ell=50-599$.  The red vertical line is the 
  observed value of 2.65, with a probability to exceed of 5.11\%.
  Middle:  histogram of
  $\max(-\mathcal{E}_\ell/\sigma(\mathcal{E}_\ell))$, which is the
  greatest significance found in any $\ell$ bin having an odd excess.  
  Although the histograms are not normal distributions, the
  observed maximum even excess is equivalent to a $\sim 1.6\sigma$ result,
  and the observed maximum odd excess of 0.34 at the vertical red line is 
  equivalent to a $\sim 2.5\sigma$ result.  Bottom:  the histogram is the same as 
  the one above, but restricted to a pre-selection of the simulations where the 
  maximum of the significance of the even excess is
  greater than or equal to the observed even excess.
\label{fig:sighist}}
\end{figure}

The results of the de-biased test are shown in Figure
\ref{fig:sighist}. In addition to the avoidance of bin selection, this
test also incorporates negative excursions of $\mathcal{E}_\ell$,
which are excesses of power at odd $\ell$.  The test shows that the
visually striking even excess in the $\ell=300-350$ bin is actually
of low significance, with a PTE of $5.1\%$ (top of Figure 
\ref{fig:sighist}).  However, the test also shows that large
excursions in odd-$\ell$ power are less frequent in the observed
power spectrum than in the Monte Carlos, such that $99.3\%$ of
the Monte Carlos have a bin with greater odd-$\ell$ significance
than the observed data (bottom of Figure 
\ref{fig:sighist}).  Thus there appears to be a modestly
significant suppression of odd-$\ell$ power.  This 
effect is only slightly relieved by accounting for the posterior 
selection of the enhanced even excess, as seen in the bottom panel of 
Figure \ref{fig:sighist}.  

We find no evidence for a radiometer-dependence of the effect.  
We were originally suspicious that the effect could arise from an interaction 
of the foreground mask with large-scale power in the map, but our simulation 
results dismissed this suspicion.

Steps and other sharp features in the power spectrum $P(k)$ tend to be 
smeared out in translation to $C_l$ space.  
For example, for the non-standard ``meandering'' cosmological inflation 
model of \cite{tye/xu:2010} 
the scalar mode responsible for inflation meanders in a multi-dimensional 
potential. This leads to a primordial CMB power spectrum with complicated 
small-amplitude variations with wavenumber $k$.  Conversion to $C_l$ 
has the effect of a significant amount of smoothing (see, for example, Figure 
1.4 of \cite{wright:2003}).  It is not likely that 
any cosmological scenario can cause the observed odd excess deficit.  
Likewise, we are aware of no experimental effects that could cause an  
odd excess deficit.  We therefore conclude that this $<3\sigma$ effect is 
most likely a statistical fluke.

\section{Aligned Quadrupole and Octupole}\label{alignedquad}

The alignment of the quadrupole and octupole was first pointed out by 
\cite{tegmark/deoliveira-costa/hamilton:2003} and later elaborated on by 
\cite{schwarz/etal:2004}, \cite{land/magueijo:2005b}, and \cite{land/magueijo:2005a}.
The fact of the alignment is not in doubt, but the significance and implications 
of the alignment are discussed here.

Do foregrounds align the quadrupole and dipole? \cite{chiang/etal:2007} conclude 
that the lowest spherical harmonic modes in the ILC map are significantly contaminated 
by foregrounds.
\cite{park/etal:2007} find that the residual foreground emission in a map resulting from 
their own independent foreground analysis is not statistically important to the large-scale 
modes of CMB anisotropy.   The large-scale modes of their map show anti-correlation with the 
Galactic foreground emission in the southern hemisphere, but they are agnostic on whether 
this is due to residual Galactic emission or by simply a matter of chance.
\cite{park/etal:2007} also assess the \iWMAP Team's ILC map and conclude that 
residual foreground emission in the ILC map does not affect the estimated large-scale values significantly.
\cite{tegmark/deoliveira-costa/hamilton:2003} also performed their own 
foreground analysis and conclude that their CMB map is clean enough that 
the lowest multipoles can be measured without any galaxy cut at all.  They 
also point out that much of the CMB power falls within the Galaxy cut region, 
``seemingly coincidentally." In other words, they conclude that the 
residual foregrounds are subdominant to the intrinsic CMB signal even 
without any Galaxy cut so long as a reduced foreground map is used. 
\cite{oliveira-costa/tegmark:2006} believe that it is more 
likely that the true alignment is degraded by foregrounds rather than created by foregrounds.

We determine the direction of the quadrupole by rotating the coordinate system until 
$|a_{\ell, \ell}|^2 + |a_{\ell, -\ell}|^2$
is maximized, where $\ell=2$ and where $a_{\ell, m}$ are the spherical harmonic coefficients.
This maximizes the power around the equator of the coordinate system.
The optimization is done by numerically checking the value
of this quantity where the $z$ vector of the coordinate system is rotated
to the center of each half-degree pixel in a res 7 ($N_{\rm side}=128$) HEALPix pixelization.
The pixel where this value is maximized is taken to be the ``direction'' 
of the quadrupole.  The pixel on the opposite side of the sphere
has the same value; we arbitrarily pick one.
Using a similar method, a ``direction'' is found for the octupole, with $\ell=3$.

\begin{figure}[htp]
\epsscale{1.0}
\plotone{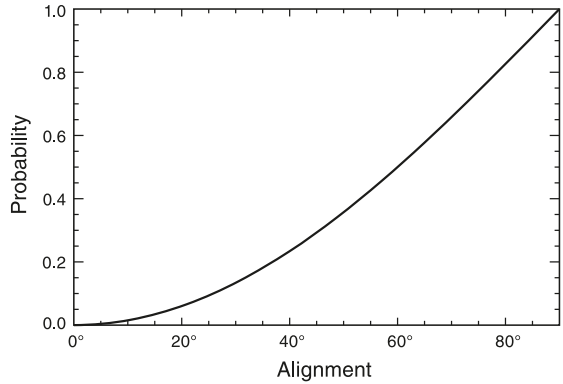}
\caption{Probability of a greater quadrupole-octupole alignment is given as a 
function of the alignment angle, in degrees.  The probability given here does not account for the
{\it a posteriori} selection of a multipole alignment search, nor does it take into account the
choice of the quadrupole and the octupole moments in particular.
\label{alignprob}}
\end{figure}

The probability that $l=2$ and $l=3$ multipoles would be aligned is shown in Figure 
\ref{alignprob}.  The $<1^\circ$ alignment in our sky appears to be quite improbable 
based upon random simulations of the best-fit $\Lambda$CDM model.  

A resolution level 7 map has 196,608 pixels of about $0\fdg5$ diameter.  
The best-fit alignment axis specifies 
two pixels directly opposite each other on the sphere.  The probability of two axes 
randomly aligning in the same pair of pixels is then 2/196,608 = 0.001\%.
The probability of getting an alignment within $0\fdg25$ of a given axis is 
0.00095\%, which is close to 0.001\% above.  

In an attempt to gain insight into the alignment we start with the ILC full-sky 
temperature map.  We then produce a map of $\Delta T^2$, which is a map of anisotropy 
power.  This map is constructed by
smoothing the seven-year ILC map by $10^\circ$, removing the mean value, and squaring.
Since the ILC map is already smoothed to $1^\circ$, the total smoothing  
corresponds to a Gaussian with FWHM = $\sqrt{10^2+1^2}^\circ = 10\fdg.05$.
We then create various masks to probe whether the dipole-quadrupole alignment can be 
attributed to one or perhaps two localized features in the map. 
The edges of some of the masks are found from contours 
of the $\Delta T^2$ map.  
The contours were selected by eye, from a gray-scale Mollweide projection in an 
image editing program, and then converted to HEALPix fits files.
Other trial masks were chosen more randomly, again to probe the sensitivity of the 
alignment to different regions of the map.

For each mask, we take the seven-year ILC map that has been smoothed by $10^\circ$,
zero the region inside the mask, and take a spherical harmonic transform.
From the $a_{\ell m}$ coefficients, we determine the angle between 
the quadrupole and octupole.

\begin{figure*}[t]
\epsscale{0.95}
\plotone{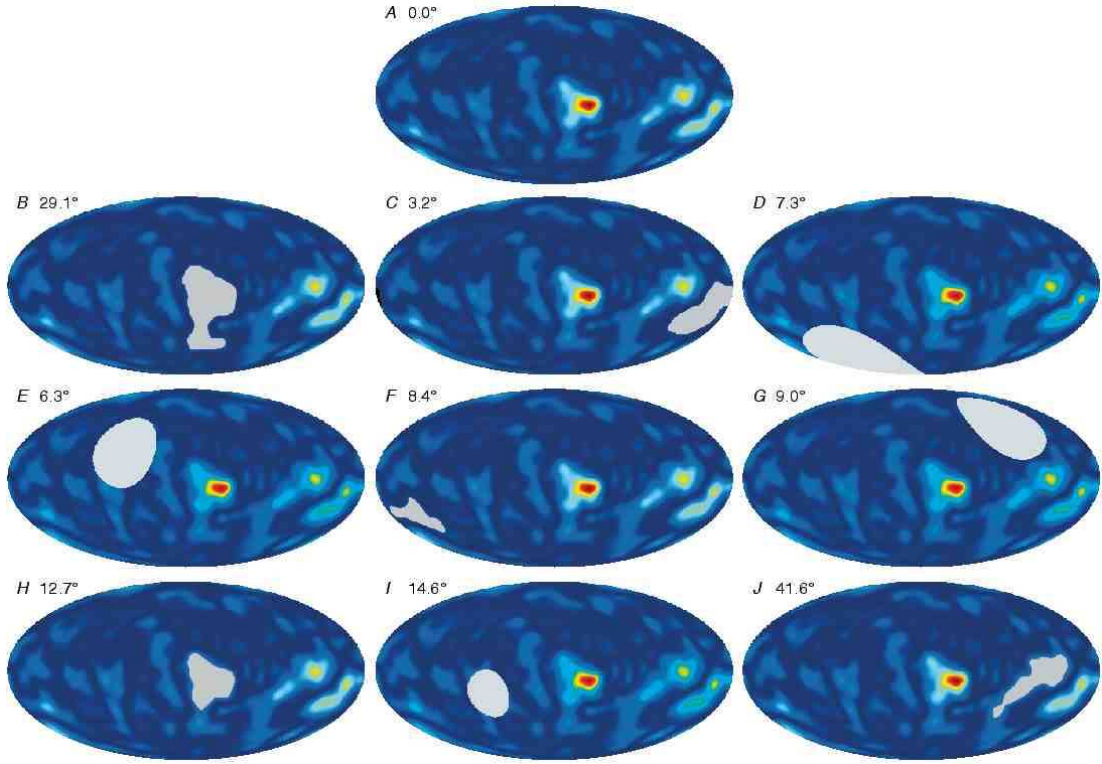}
\caption{We smooth and square the ILC map, as described in the text, to make a map of 
$(\Delta T)^2$ to roughly visualize the anisotropy power in a sky map (A).  
Cold Spot I is the highest power region at this level of resolution, but 
several other regions also contribute substantial power at this scale.  
For map (A), the quadrupole and octupole are aligned to within $\sim 1^\circ$. 
To evaluate what regions contribute to this alignment, we mask selected gray areas as shown 
in the sky maps (B) through (J).  Each map is labeled by the degree of quadrupole-octupole 
alignment remaining after the gray mask is applied.  Masking Cold Spot I in (B) or (H) 
eliminates any significant alignment.  However, keeping those regions and masking  
other regions, also breaks the alignment to a significant degree.  We conclude that no 
single region or pair of regions generates the alignment.  Rather, the
combined power contributions over
a substantial fraction of the full-sky map cause the high degree of 
quadrupole--octupole alignment.  Note that the chance alignment of CMB power with the 
Galactic cut region discussed in Section \ref{largescalepower} is apparent in map (A).
\label{maskT2}}
\end{figure*}

\begin{figure}[htp]
\epsscale{1.0}
\plotone{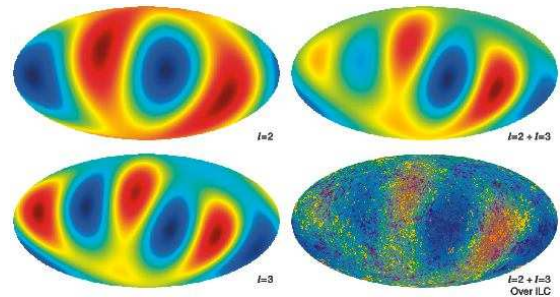}
\caption{$l=2$ quadrupole and $l=3$ octupole maps are added.  The combined map is
then shown superposed on the ILC map from Figure \ref{fingers}.  Note that the quadrupole 
and octupole components arrange themselves to match the cool fingers and the warm regions
in between.  The fingers and the alignment of the $l=2$ and $l=3$ multipoles are intimately 
connected.
\label{multipole}}
\end{figure}

Figure \ref{maskT2} shows the smoothed, squared
temperature map (in color) and the effect of various masks (in gray) 
on the quadrupole--octupole alignment.  
Masking Cold Spot I eliminates any significant alignment.  
However, keeping that region but masking other regions also significantly 
reduces the quadrupole--octupole alignment.   The posterior selection of the 
particular masked regions is irrelevant as the point is only to demonstrate that no 
single region or pair of regions solely generates the $<1^\circ$ alignment.  Rather the 
high degree of quadrupole--octupole alignment results from the statistical distribution of
anisotropy power over the whole sky. This rules out single-void 
models, a topological defect at some sky position, or any other such explanation.  
The alignment behaves as one would expect if it originates from chance random 
anisotropy amplitudes and phases.  The alignment of the $l=2$ and $l=3$ multipoles 
is intimately connected with the large-scale cool fingers and intervening warm regions, 
discussed earlier, as can be seen in Figure \ref{multipole}.   
Although the alignment is indeed remarkable, 
current evidence is more compatible with a statistical combination of full-sky data 
than with the dominance of one or two discrete regions.

\cite{francis/peacock:2010} estimated the local ($z<0.3$) density field from the 
2MASS and SuperCOSMOS galaxy catalogs and used that field to estimate the
ISW effect within this volume.  Large-scale features were extrapolated 
across the Galactic plane.  The effects of radial smearing from the photometric galaxy 
sample were reduced by taking only three thick redshift shells with $\Delta z =0.1$.  A 
linear bias was used to relate galaxies to density, independent of both scale and redshift 
within each of the three shells.  They estimated that the $z<0.3$ data limit contains 
$\sim 40$\% of the total ISW signal.  \cite{francis/peacock:2010} removed their estimate of the 
ISW effect from the \iWMAP map.  One result was to raise the amplitude of the quadrupole while 
the octupole amplitude was relatively unchanged.  More importantly, they reported that there 
remains no significant quadrupole--octupole alignment after the ISW removal.  With the  
\cite{francis/peacock:2010} result, the quadrupole--octupole alignment shifts from an
early universe property to a statistical fluke that the secondary anisotropy 
effect from the local density distribution happens to superpose on the primordial anisotropy 
in such a way as to align the quadrupole and octupole.  

\section{Hemispherical and Dipole Power Asymmetry}\label{hemispherical}

Claims of a dipole or hemispherical power asymmetry in \iWMAP\ maps have appeared in the literature
since the release of the first-year \iWMAP\ data, with estimates of statistical significance 
ranging up to $3.8\sigma$. We distinguish between a ``hemispherical'' power asymmetry, in which 
the power spectrum is assumed to change discontinuously across a great circle on the sky, and a 
``dipole'' power asymmetry in which the CMB is assumed modulated by a smooth 
cosine function across the sky, i.e., the CMB is assumed to be of the form
\begin{equation}
T({\bf n})_{\rm modulated} = \left( 1 + {\bf w} \cdot {\bf n} \right) T({\bf n})_{\rm unmodulated}.  
\label{eq:dipole_modulation}
\end{equation}
Previous analyses of \iWMAP data in the literature have fit for either hemispherical or dipolar
power asymmetry, 
and the results are qualitatively similar: asymmetry is found with similar 
direction and amplitude in the two cases.  However, analyses that use optimal estimators (as we 
do in our analysis below) have all studied the dipolar modulation
\citep[e.g.,][]{hoftuft/etal:2009, hanson/lewis:2009}.  
Furthermore, theoretical attempts to obtain cosmological power asymmetry by altering the 
statistics of the primordial fluctuations \citep{gordon:2007,  donogue/dutta/ross:2009,  
erickcek/kamionkowski/carroll:2008, erickcek/etal:2009} 
have all found a dipolar modulation rather than a hemispherical modulation.  Therefore, we will 
concentrate on the dipolar modulation, defined by Equation~(\ref{eq:dipole_modulation}), 
for the sake of better comparison with 
both early universe models, and with similar analyses in the literature.  Unambiguous evidence for 
power asymmetry would have profound implications for cosmology.

We revisit this analysis and conclude that this claimed anomaly is not statistically significant, 
after {\it a posteriori} choices are carefully 
removed from the analysis. In looking for a power asymmetry, the most significant issue
is removing an arbitrary choice of scale, either specified explicitly by a maximum multipole $l$, 
or implicitly by a sequence of operations such as smoothing and adding extra noise that define 
a weighting in $l$. 

The first claimed hemispherical power asymmetry appeared
in \cite{eriksen/etal:2004}, based on the first-year \iWMAP\ data, 
where a statistic for power asymmetry was constructed,
and its value on high-resolution \iWMAP\ data was compared to
an ensemble of Monte Carlo simulations in a direct frequentist approach.  They quoted a 
statistical significance of 95\%--99\%, depending on the range of $l$ selected.  
The details of this analysis contained many arbitrary choices.  
\cite{hansen/etal:2004, hansen/etal:2009} also used a similar methodology. 
In \cite{hansen/etal:2009}, the significance of a $2 \le l\le 600$ hemispherical power 
asymmetry was quoted as 99.6\%.

A second class of papers used a low-resolution pixel likelihood formalism to study power
asymmetry.
\cite{eriksen/etal:2007c} used this
approach to search for a dipole power asymmetry in low-resolution 
three-year \iWMAP\ data and a statistical significance of $\sim 99$\% was claimed.
\cite{hoftuft/etal:2009} repeated the likelihood analysis at somewhat
higher resolution and quoted a statistical significance of 3.5$\sigma$ -- 3.8$\sigma$
for different choices of resolution.
Although the likelihood estimator contained fewer arbitrary choices than the preceding
class of papers, the low-resolution framework still contained an arbitrary choice of
angular scale, which may be tuned (intentionally
or unintentionally) to spuriously increase statistical significance.
\cite{hoftuft/etal:2009} introduced an explicit cutoff multipole $l_{\rm mod}$
and the CMB signal was assumed to be unmodulated for $l>l_{\rm mod}$ and modulated
for $l\le l_{\rm mod}$.
Both \cite{eriksen/etal:2007c} and \cite{hoftuft/etal:2009} downgraded the data in
resolution, smoothed with a Gaussian window, and added extra white noise.
These processing steps implicitly defined a weighting in $l$ where the power asymmetry 
is estimated, and introduced arbitrary choices into the analysis.

A third approach to the analysis, based on optimal quadratic estimators,
recently appeared in \cite{hanson/lewis:2009}.
In this approach, the \iWMAP\ data were kept at full resolution and a minimum-variance
quadratic estimator was constructed for each of the three vector components of the 
dipole modulation ${\bf w}_i$.
Hanson and Lewis found $\approx 97$\% evidence for a dipole power asymmetry
at $2\le l \le 40$, and $\approx 99.6$\% evidence for dipole power asymmetry at $2\le l\le 60$. 
However, the result was strongly dependent on changing the $l$ range, and quickly went away for 
higher $l$. A significant shift was seen between the KQ75 and KQ85 masks, and between raw and 
clean maps, suggesting that foreground contamination was not negligible.

\begin{figure}[htp]
\epsscale{1.0}
\plotone{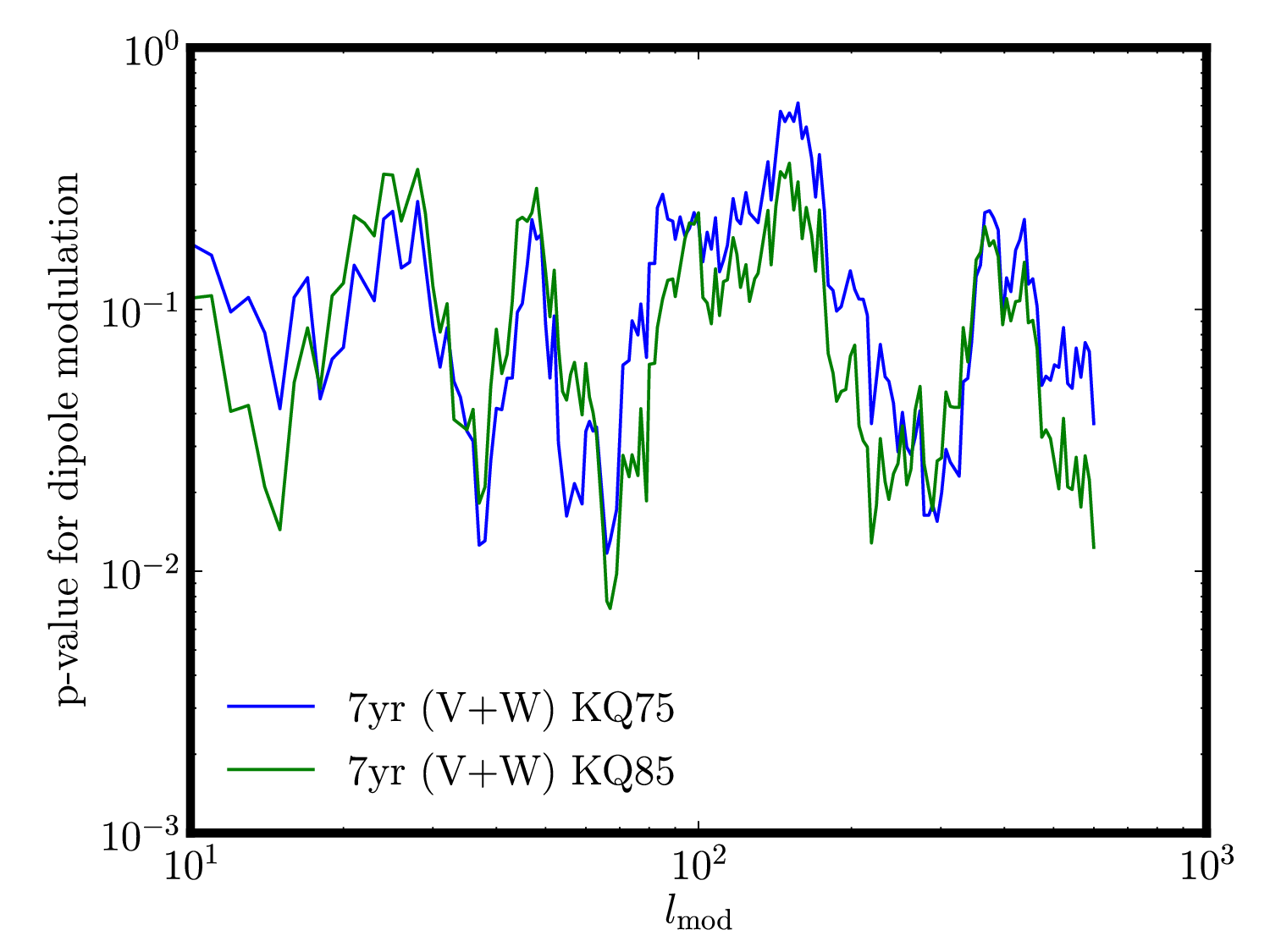}
\caption{Probability for a Monte Carlo simulation to have a larger dipole modulation 
than the co-added V+W \iWMAP data, as measured by the statistic $\kappa_1$, is shown as a 
function of the maximum multipole moment $l_{\rm mod}$ that is assumed to be modulated.  
This can be interpreted as the statistical significance of power asymmetry for a fixed 
value of $l_{\rm mod}$, if one does not account for possible {\it a posteriori} bias when 
choosing $l_{\rm mod}$.  
\label{dm}}
\end{figure}

Comparing these methods, we find that the \cite{hanson/lewis:2009} optimal quadratic estimator 
has significant advantages over other analysis methods that have appeared in the literature.
\begin{enumerate}
\item There are no arbitrary choices (such as smoothing scale) in the optimal quadratic estimator.
One can either look for power asymmetry in a range of multipoles $2\le l \le l_{\rm mod}$, or over
the entire range of angular scales measured by \iWMAP, and the estimator is uniquely determined by the
minimum-variance requirement in each case.  None of the previously considered methods had this property.
\item There is no need to degrade the \iWMAP\ data, or include processing steps such as adding extra
noise, since the optimal quadratic estimator can be efficiently
computed at full \iWMAP\ resolution using the multigrid $C^{-1}$ code from \cite{smith/zahn/dore:2007}.
\item Statistical significance can be assessed straightforwardly by comparing the estimator with an
ensemble of Monte Carlo simulations.  In particular, maximum likelihood analyses in the literature
have assessed significance using Bayesian evidence, but schemes for converting the evidence integral
into a frequentist probability are not a sufficient substitute for true Monte Carlo simulations, 
which directly give the probability for a simulation to be as anomalous as the data.
\end{enumerate}
For these reasons, we have studied dipole power asymmetry using the optimal quadratic estimator.
We introduce a cutoff multipole $l_{\rm mod}$, and assume that the CMB is isotropic for $l > l_{\rm mod}$
and dipole-modulated for $2 \le l \le l_{\rm mod}$.
There is an optimal quadratic estimator ${\hat w}_i$ for each of the three components of the 
(vector) modulation ${\bf w}$ (Equation~(\ref{eq:dipole_modulation})), and an estimator $\hat\kappa_1$ for the 
(scalar) amplitude of the modulation.
Implementation details of the estimators are presented in the Appendix, 
where we also comment on the relation with maximum likelihood.

Figure \ref{dm} shows the probability that the value of the dipole modulation statistic 
$\hat\kappa_1$ is larger than for the \iWMAP\ data, when evaluated by Monte Carlo simulation.
There are choices of $l_{\rm mod}$ where the power asymmetry appears to have high 
significance. 
For example, when we chose the KQ85y7 mask and $l_{\rm mod} = 67$, the probability for a 
simulation to have a
value of $\hat\kappa_1$ that is larger than the data is 0.7\%.
This could be interpreted as 2.5$\sigma$ evidence for a power asymmetry, but such an 
interpretation would be inflating the statistical significance since the choice of 
$l_{\rm mod}$ is an {\it a posteriori} choice. Consider an analogous example for the 
five-year $C_l$ power spectrum.  The power in $C_{l=512}$ is high by $3.7\sigma$, but this 
is not really a $3.7\sigma$ anomaly.  Rather, it reflects the fact that there are a large 
number of $l$ values that could have been chosen.

Now we seek to assess the global statistical significance of the power asymmetry without 
making any {\it a posteriori} choices.  Consider the probability for a Monte Carlo simulation to
have a larger value of $\hat\kappa_1$ than the \iWMAP\ data, as a function 
of $l_{\rm mod}$.  This can be interpreted as the statistical significance for power asymmetry
in the range $2\le l\le l_{\rm mod}$, for a fixed choice of $l_{\rm mod}$. 
Let $\eta$ be the minimum value of the probability, which we find to be
$\eta=0.012$ with the KQ75y7 mask (corresponding to $l_{\rm mod}=66$), or $\eta=0.007$ with 
the KQ85y7
mask (corresponding to $l_{\rm mod}=67$). We now assess whether this value of
$\eta$ is anomalously low.  To determine this, we compute $\eta$ for an ensemble of Monte 
Carlo simulations where $l_{\rm mod}$ is chosen to maximize the value of $\eta$ 
independently for each simulation.

We perform the maximization over the range $10 \le l_{\rm mod} \le 132$.  (The results depend 
only weakly on this choice of range; we have taken the upper limit of the range to be 
twice as large as the most anomalous $l_{\rm mod}$ in the \iWMAP data.)  
We find that the probability that a 
simulation has a value of $\eta$ that is smaller than for \iWMAP\ is 13\% for the KQ75y7 mask,
and 10\% for the KQ85y7 mask.  

Motivated by the power asymmetry, \cite{erickcek/etal:2009} presented a variation of the 
curvaton inflationary scenario in which the curvaton has a large-amplitude super-horizon 
spatial gradient that modulates the amplitude of CMB anisotropy, thereby generating a 
hemispherical power asymmetry that could match the CMB data.  \cite{hirata:2009} used 
high-redshift quasars to place a limit on the gradient in the amplitude of perturbations 
that would be caused in this scenario.  Their limit ruled out the simplest version of 
this curvaton spatial gradient scenario.  Our new CMB results, presented here, largely 
remove the initial motivation for this theory.

We conclude that there is no significant evidence for an 
anomalous dipole power asymmetry in the \iWMAP\ data.

\section{Quadrupolar Dependence of the Two-Point Function}\label{quadrupolar}

There is another effect, related to the dipole power asymmetry from
Section \ref{hemispherical},
in which the two-point function of the CMB contains a component that varies as a quadrupole on the
sky.  Motivated by an anisotropic model of the early universe proposed by \cite{ackerman/etal:2007}
that predicts such a signal, \cite{groeneboom/eriksen:2009} used a Gibbs sampler to claim
``tentative evidence for a preferred direction'' of $(l,b)=(110^\circ ,10^\circ)$  
in the five-year {\it WMAP} map. A theoretical model that predicts large-scale quadrupolar 
anisotropy was also proposed by \cite{gordon/etal:2005}. 
A similar effect could be caused by {\it WMAP}'s 
asymmetric beams, which were not accurately represented in this work, and an algebraic
factor was missing in the analysis.  
In \cite{hanson/lewis:2009}, the missing algebraic factor was corrected and the effect 
was verified with high statistical significance, using an optimal cut-sky quadratic estimator.
Optimal estimators had previously been constructed in the all-sky case by 
\cite{pullen/kamionkowski:2007} and \cite{dvorkin/peiris/hu:2008}.

Recently, \cite{groeneboom/etal:2010} returned another fit, this time including polarization, 
the factor correction, examinations of beam asymmetries, noise misestimation, 
and zodiacal dust emission.  The new claim was 9$\sigma$ evidence of the preferred direction 
$(l,b)=(96^\circ ,30^\circ)$, which was 
quite far from the original alignment direction claimed.  The new preferred direction 
was toward the ecliptic poles, strongly suggesting that this is not a 
cosmological effect.  The claimed amplitude was frequency dependent, also inconsistent 
with a cosmological effect.  Zodiacal dust emission 
was ruled out as the source of the alignment.  \cite{hanson/lewis:2009}
found that the beam asymmetry was a large enough effect to explain the signal, although 
Groeneboom et al. reported the opposite conclusion. 
The claimed statistical significance of the quadrupolar power asymmetry is so high that it seems 
impossible for it to be a statistical fluke or built up by posterior choices, even 
given the number of possible anomalies that could have been searched for. 

We have implemented the optimal quadratic estimator following the approach
of \cite{hanson/lewis:2009} and confirmed that the effect exists with high statistical significance.
Rather than presenting an analysis that is tied to a particular model (either cosmological or
instrumental), we have found it convenient to parameterize the most general
quadrupolar power asymmetry using the language of the bipolar spectrum from \cite{hajian/souradeep:2003}.
This is reviewed in the Appendix. The summary is that the most general quadrupolar anomaly 
can be parameterized by two quantities $A_{ll}^{2M}$ and $A_{l-2,l}^{2M}$ which
are $l$-dependent and have five components corresponding to the degrees of freedom of a
quadrupole.
If statistical isotropy holds, then $A_{ll}^{2M} = A_{l-2,l}^{2M} = 0$.
The special case where $A_{ll}^{2M} \approx A_{l-2,l}^{2M} \ne 0$ corresponds to an
anisotropic model in which the local power spectrum varies across the
sky (i.e., the quadrupolar 
analogue of the dipole modulation in
Equation~(\ref{eq:dipole_modulation})).
The special case where $A_{ll}^{2M} \approx -2 A_{l-2,l}^{2M} \ne 0$ corresponds to 
an anisotropic model in which the local power spectrum is isotropic, but hot and cold spots 
have preferred ellipticity where the local magnitude and orientation varies across the sky.
Thus there are two independent ``flavors'' of quadrupolar anomaly; the bipolar power spectrum
distinguishes the two and also keeps track of the $l$ dependence.
A proposed model for the quadrupolar effect in the \iWMAP data can be tested by
computing the bipolar power spectrum of the model, and comparing with estimates of
the bipolar spectrum from data.

\begin{figure}[htp]
\epsscale{1.0}
\plotone{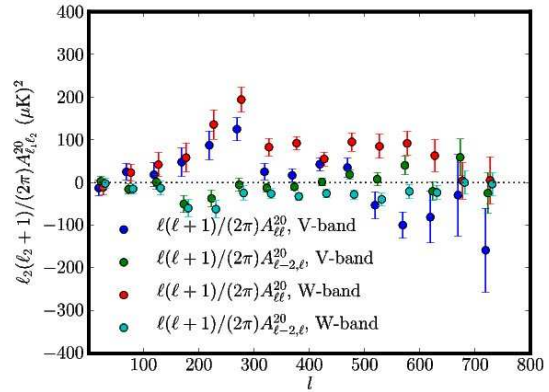}
\caption{Quadrupolar bipolar power spectra, binned with $\Delta l = 50$, are 
shown separately for V-band and W-band \iWMAP data, using the KQ75y7 mask.  Only 
the components of the bipolar power spectra that 
point along the ecliptic axis are shown (i.e., components with $M=0$ in ecliptic coordinates).  
A statistically significant quadrupolar effect is seen, even for a single frequency band 
in a single angular bin.
\label{qm}}
\end{figure}

Figure \ref{qm} shows the components of the bipolar power spectrum of the \iWMAP\ data
that point along the ecliptic axis (i.e., $A_{ll}^{20}$ and $A_{l-2,l}^{20}$ in ecliptic 
coordinates).
A nonzero bipolar power spectrum is seen with high statistical significance, even in a single bin
with $\Delta l=50$, confirming the existence of a quadrupolar effect.

We implemented a number of diagnostic tests to characterize the quadrupolar effect; our findings can be
summarized as follows:
\begin{enumerate}
\item Only the components of the bipolar power spectrum that point in the ecliptic direction
(i.e., components $A^{2M}_{l_1 l_2}$ with $M=0$ in ecliptic coordinates) contain a statistically significant
signal.  The components with $M=1$ or $M=2$ are consistent with zero within their statistical errors,
even if we sum over all values of $l$ to maximize signal-to-noise.
\item The effect is larger in W-band than V-band, which is inconsistent with a cosmological origin.
\item The angular dependence of the effect shows a bump at the scale of the first acoustic peak
($l\approx 220$), disfavoring an explanation from foregrounds or noise, which would
not be expected to show acoustic peaks.
\item If we split the optimal quadratic estimator into contributions from cross correlations between
differential assemblies (DA's) in \iWMAP, and auto correlations in which each DA is correlated with
itself, then we find that the amplitude of the effect is consistent in the two cases, disfavoring
instrumental explanations that are not highly correlated between channels (such as striping due
to $1/f$ noise).
\item The bipolar power spectrum of \iWMAP satisfies $A_{\ell\ell}^{20} \approx -2A_{\ell-2,\ell}^{20}$,
corresponding to a model in which the small-scale power spectrum is isotropic, but the shapes of hot
and cold spots are not.  (In fact, for this reason, we have used the term ``quadrupolar effect'' in
this section rather than ``quadrupolar power asymmetry'', which would suggest that the power spectrum
is modulated.  We favor the label ``effect'' over ``anomaly'' because it is only an anomaly in the
absence of a plausible cause.)
\end{enumerate}

Given the strong ecliptic alignment and that the ecliptic plane was the symmetry axis of the \iWMAP observations, 
and the non-blackbody frequency dependence, we conclude that this is not a $\Lambda$CDM anomaly.  
It seems very likely that the observed quadrupolar effect is the result of incomplete handling of beam asymmetries.
Beam asymmetry generates an instrumental bipolar power spectrum that
is consistent with all five items above,
and it is difficult to think of any other instrumental contribution that satisfies these properties.
However, we have not yet simulated the effects of asymmetric beams to confirm this explanation.
A full investigation of the effect of beam asymmetry is underway and preliminary indications from 
our work to date are consistent with our hypothesis.

While a detailed explanation of the quadrupolar effect is pending, it is important to have as much 
confidence as possible
that a large anomaly in \iWMAP does not bias the estimated power spectrum.
It is reassuring (item 5 above) that the angle-averaged power spectrum appears to be statistically 
isotropic,
suggesting that the power spectrum is ``blind'' to the effect (or, less sensitive to beam
asymmetries, assuming that is the cause).
Furthermore, if beam asymmetry does turn out to explain the quadrupolar effect, 
then the analysis in Appendix B
of \cite{hinshaw/etal:2007} shows independently that the power spectrum bias due to beam asymmetry is small.
Nevertheless, it is important to follow up on the studies to date, and we plan to do so in the future.

\section{Conclusions}\label{conclusions}
In the context of this paper we take an ``anomaly'' to refer to a statistically unacceptable 
fit of the $\Lambda$CDM model to the $C_l$ data, a statistically significant deviation of the $a_{lm}$ 
from Gaussian random phases, or correlations between the $a_{lm}$.   We are not concerned here 
with the current uncertainty range of parameter values allowed by the $\Lambda$CDM model or with 
whether an alternative model is also an acceptable fit to the data.  

Numerous claims of \iWMAP\ CMB anomalies have been published.  We find that there are a few  
valuable principles to apply to assess the significance of suspected anomalies: (1) 
human eyes and brains are excellent at detecting visual patterns, but poor at assessing 
probabilities.  Features seen in the \iWMAP\ maps, such as the large Cold Spot I near the 
Galactic center region, can stand out as unusual.  However, the likelihood of such features 
cannot be discerned by visual inspection of our particular realization of the universe. 
(2) Monte Carlo simulations are an invaluable way to determine the expected deviations 
within the $\Lambda$CDM model.  Claims of anomalies without Monte Carlo simulations are 
necessarily weak claims. (3) Some parameters are weak discriminants of cosmology because 
they take on a broad 
range of values for multiple realizations of the same model.  (4)  {\it A posteriori} 
choices can have a substantial effect on the estimated significance of 
features.  For example, it is not unexpected to find a $2\sigma$ feature when analyzing a rich 
data set in a number of different ways.  However, to assess whether a particular $2\sigma$ 
feature is interesting, one is often tempted to narrow in on it to isolate its behavior.  
That process involves {\it a posteriori} choices that amplify the apparent significance of 
the feature.  

\begin{figure}[htp]
\epsscale{1.0}
\plotone{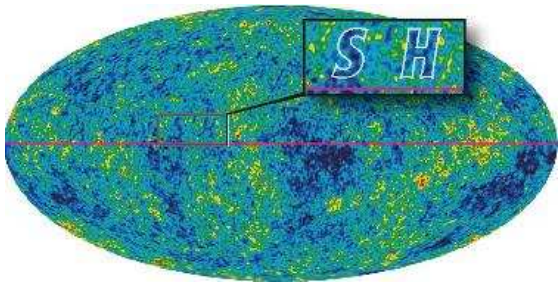}
\caption{``SH'' initials of Stephen Hawking are shown in the ILC sky map.  The ``S'' and 
``H'' are in roughly the same font size and style, and both letters are aligned neatly along a line 
of fixed Galactic latitude.  A calculation would show that the probability of this particular 
occurrence is vanishingly small.  Yet, there is no case to made for a non-standard cosmology despite 
this extraordinarily low probability event.  It is clear that the combined selection of looking 
for initials, these particular initials, and their alignment and location are all {\it a posteriori} 
choices.  For a rich data set, as is the case with \iWMAP, there are a lot of data and a lot of ways
of analyzing the data.  Low probability events are guaranteed to occur.  The {\it a posteriori} 
assignment of a likelihood for a particular event detected, especially when the detection of 
that event is ``optimized'' for maximum effect by analysis choices, does not result in a 
fair unbiased assessment.  This is a recurrent issue with CMB data analysis and is often a 
tricky issue and one that is difficult to overcome.
\label{sh}}
\end{figure}

Shortly after the \iWMAP\ sky maps became available, one of the authors (L.P.) noted that 
the initials of 
Stephen Hawking appear in the temperature map, as seen in Figure \ref{sh}.   Both the ``S'' and ``H" are beautifully vertical in Galactic coordinates, 
spaced consistently just above the $b=0$ line.  We pose the question, what is the probability of 
this occurrence?  It is certainly infinitesimal; in fact, much less likely than several 
claimed cosmological anomalies.  Yet, we do not take this anomaly seriously because it is silly.  
The Stephen Hawking initials highlight the problem with {\it a posteriori} statistics.  By looking 
at a rich data set in multiple different ways, unlikely events are expected.  The search for 
statistical oddities must be viewed differently from tests of pre-determined hypotheses.  The data 
have the power to support hypothesis testing rooted in ideas that are independent of the \iWMAP
data.  We can ask which of two well-posed theoretical ideas is best supported by the data.  Much of
the \iWMAP\ analysis happens in a different context asking, ``What oddities can I find in the data?"
 
For example, no one had predicted that low-l multipoles might be aligned.  Rather, this followed
from looking into the statistical properties of the maps.
Simulations, both by the \iWMAP\ team and 
others, agree that this is a highly unusual occurrence for the standard $\Lambda$CDM cosmology.  
Yet, a large fraction of simulated skies will likely have some kind of oddity.  The key is 
whether the oddity is specified in advance.  

The search for oddities in the data is essential for testing the model. The success of the model
makes these searches even more important.  A detection of any highly significant {\it a posteriori} 
feature could become a serious challenge for the model.  The less significant features discussed 
in this paper provided the motivation for considering alternative models and developing new 
analysis of {\it WMAP} (and soon {\it Planck}) data.  The oddities have triggered proposed new 
observations that can further test the models.

It is often difficult to assess the statistical claims. It may well be that an oddity could be
found that motivates a new theory, which then could be tested as a hypothesis against 
$\Lambda$CDM. 
The data support these comparisons.  Of course, other cosmological measurements must also play 
a role in testing new hypotheses.
No CMB anomaly reported to date has caused the scientific community to adopt 
a new standard model of cosmology, but claimed anomalies have been used to provoke thought 
and to search for improved theories.

We find that Cold Spot I does not result from Galactic foregrounds, but rather forms the 
northernmost part of one of four cool ``fingers'' in the southern sky.  Its amplitude 
and extent are not unusual for $\Lambda$CDM.  In fact, structures with this nature are expected.  

We find that Cold Spot II is at the southernmost end of a different one of the southern 
fingers, and it has been shown not to be an anomalous fluctuation.

We find that the amplitude of the $l=2$ quadrupole component is not anomalously low, but well 
within the 95\% confidence range.  

We conclude that there is no lack of large-scale CMB power over the full \iWMAP sky.
The low value of the $S$-statistic integral over the large-angle correlation function 
on the cut sky results from a posterior choice of a sub-optimal (i.e., not full sky) 
statistic, $S_{1/2}$, a chance alignment of the Galactic plane cut with CMB signal, 
and a chance alignment of primary CMB fluctuation features with secondary ISW features 
from the local density distribution.

We find that the quadrupole and octupole are aligned to a remarkable degree, but that 
this alignment is not due to a single feature in the map or even a pair of features.
The alignment does not appear to be due to a void, for example.  
We find that the alignment is intimately associated with the fingers of 
the large-scale anisotropy visible in the southern sky, and it results from the statistical 
combination of fluctuations over the full sky.  
There is also evidence that the alignment is due, in part, to a coincidental alignment 
of the primary anisotropy with the secondary anisotropy from the local density distribution 
through the ISW effect.  At the present time the remarkable degree of alignment 
appears to be no more than a chance occurrence, discovered {\it a posteriori} with no 
motivating theory.  A new compelling theory could change this conclusion.
 
There is a portion of the power spectrum where there is a marginally significant lack 
of odd multipole power relative to even multipole power, but overall the \iWMAP data 
are well fit by the $\Lambda$CDM model.  There is no systematic error that we are aware of 
that could cause the even power excess, nor are there any cosmological effects that
would do so.  We conclude that the even excess is likely a statistical fluctuation that 
was found {\it a posteriori}.  No motivating theory for this phenomenon is known.

We find that claims of hemispherical and/or dipole asymmetries have suffered from {\it a
posteriori} choices.   After carrying out an analysis in a manner that avoids {\it a
posteriori} bias, we find that the evidence for a hemispherical power asymmetry is weak.  

Evidence has been reported for a significant quadrupolar power asymmetry that does 
not appear to be cosmological in origin, and most likely results from an incomplete 
propagation of beam asymmetries.  A careful analysis will be a subject of future work.

We have examined selected claims of CMB anomalies, but this paper is not a
comprehensive review article and we have not attempted to address every anomaly
paper in the literature. However, we can extend our results by recognizing that
various claims of anomalies are not necessarily independent from those
that have been
examined. For example, there are smaller-scale consequences of the fact that the
value of the quadrupole on our sky is low (but not anomalously low) compared
with the maximum likelihood expected value for the best-fit $\Lambda$CDM model.
The amplitudes of moderate to small scale hot and cold spots are expected to be
reduced statistically relative to their predicted  amplitude in the best-fit
$\Lambda$CDM model due to the reduced contribution from the quadrupole
component. That is, the temperature anisotropy in a specific spot has
contributions from a range of multipoles, including the quadrupole contribution.
Since the power in the multipoles scales as $l^{-2}$, the low quadrupole value
in our sky statistically reduces the amplitude stretch of hot and cold spots in
the map of our sky. This should help, at least in part, to explain the results
of \cite{hou/etal:2009}, \cite{monteserin/etal:2008}, \cite{ayaita/etal:2010},
and \cite{larson/wandelt:2004}.  Another example where the results from this
paper may relate to a claimed anomaly is our report of a quadrupolar power
asymmetry. This may be the effect that was detected by \cite{wiaux/etal:2006}
through an analysis with a second gaussian derivative.

\acknowledgments
The \iWMAP mission is made possible by the support of the NASA Science Mission Directorate.  
This research has made use of NASA's Astrophysics Data System 
Bibliographic Services. We thank Jon Urrestilla for sharing the texture power spectrum 
with us. 
We also acknowledge use of the HEALPix \citep{gorski/etal:2005},  
CAMB \citep{lewis/challinor/lasenby:2000}, and CMBFAST \citep{seljak/zaldarriaga:1996} packages.
We are very grateful to Duncan Hanson for useful discussions.

\appendix

\section{Estimators for dipolar and quadrupolar anomalies}

In this appendix, we present the detailed construction of the estimators used to study dipole
power asymmetry in Section \ref{hemispherical} and quadrupolar dependence of the two-point function
in Section \ref{quadrupolar}.

\subsection{Bipolar power spectrum}

If statistical isotropy is assumed, then the two-point function of the CMB is parameterized 
by the power spectrum $C_\ell$: 
\begin{equation}
\langle a_{\ell_1 m_1} a_{\ell_2 m_2} \rangle = (-1)^{m_1} C_{\ell_1} \delta_{\ell_1\ell_2} \delta_{m_1,-m_2}
\end{equation}
The bipolar power spectrum, introduced in \cite{hajian/souradeep:2003}, is a formalism for analogously parameterizing the two-point function
if the assumption of statistical isotropy is relaxed.
If we decompose the two-point function $\langle a_{\ell_1 m_1} a_{\ell_2 m_2} \rangle$ into a sum of terms which transform
under rotations with total angular momentum $L$, then we arrive at the following expansion:
\begin{equation}
\langle a_{\ell_1 m_1} a_{\ell_2 m_2} \rangle 
  = \sqrt{(2\ell_1+1)(2\ell_2+1)} \sum_{LM} A_{\ell_1\ell_2}^{LM*} \sqrt{2L+1}
    \threej{\ell_1}{\ell_2}{L}{0}{0}{0} \threej{\ell_1}{\ell_2}{L}{m_1}{m_2}{M}  \label{eq:bipolar_ps_def}
\end{equation}
This equation defines the bipolar power spectrum $A_{\ell_1\ell_2}^{LM}$.
(Note that our normalization and sign convention differ from \cite{hajian/souradeep:2003}; this will simplify 
some of the equations that follow.)

The bipolar power spectrum has the following properties:
\begin{enumerate}
\item $A_{\ell_1\ell_2}^{LM}$ vanishes unless $-L \le M \le L$,
the triple $(\ell_1,\ell_2,L)$ satisfies the triangle inequality,
and $(\ell_1+\ell_2+L)$ is even.
\item Under rotations, $A_{\ell_1\ell_2}^{LM}$ transforms as a spin-$L$ object; under the parity operation $\n\rightarrow (-\n)$,
it transforms as $A_{\ell_1\ell_2}^{LM} \rightarrow (-1)^L A_{\ell_1\ell_2}^{LM}$.
\item Symmetry: $A_{\ell_2\ell_1}^{LM} = A_{\ell_1\ell_2}^{LM}$.
\item Reality: $A_{\ell_1\ell_2}^{LM*} = (-1)^M A_{\ell_1\ell_2}^{L,-M}$.
\end{enumerate}

To get some intuition for the bipolar power spectrum, we now consider a series of increasingly complicated models
and compute the bipolar power spectrum in each case.

Our first trivial example will be an isotropic model with power spectrum $C_\ell$.
In this case comparison with Equation~(\ref{eq:bipolar_ps_def}) shows that the bipolar power spectrum is given by
\begin{equation}
A_{\ell_1\ell_2}^{LM} = C_{\ell_1} \delta_{\ell_1\ell_2} \delta_{L0} \delta_{M0}
\end{equation}
In general, the $L=0$ component $A_{\ell\ell}^{00}$ of the bipolar power spectrum is equal to the
angle-averaged power spectrum $C_\ell$.
(Note that the properties above imply that the only component of the bipolar power spectrum with
$L=0$ is $A_{\ell\ell}^{00}$.)
Components with $L>0$ will parameterize deviations from statistical isotropy.

Our next example (from Section \ref{hemispherical})
is an anisotropic model which is obtained by applying a dipolar ``sky'' modulation
to an isotropic CMB with power spectrum $C_\ell$:
\begin{equation}
T(\n) = \left(1 + \sum_{M=-1}^1 w_{1M} Y_{1M}(\n) \right) T(\n)_{\rm iso}
\end{equation}
where $T_{\rm iso}(\n)$ is an isotropic CMB realization.
To first order in the modulation $w_{1M}$, a short calculation shows that the bipolar power spectrum is given by
\begin{eqnarray}
A_{\ell\ell}^{00} &=& C_\ell  \\
A_{\ell-1,\ell}^{1M} = A_{\ell,\ell-1}^{1M} &=& \frac{w_{1M} (C_{\ell-1} + C_{\ell})}{(4\pi)^{1/2}}
\end{eqnarray}
with all other components zero.
The modulation does not change the sky-averaged power spectrum $A_{\ell\ell}^{00}$, but the power spectrum in
local patches near the two poles will be different.

A very similar example is the quadrupolar ``sky'' modulation:
\begin{equation}
T(\n) = \left(1 + \sum_{M=-2}^2 w_{2M} Y_{2M}(\n) \right) T(\n)_{\rm iso}
\end{equation}
with bipolar power spectrum given by:
\begin{eqnarray}
A_{\ell\ell}^{00} &=& C_\ell  \\
A_{\ell\ell}^{2M} &=& \frac{w_{2M} C_\ell}{\pi^{1/2}}  \\
A_{\ell-2,\ell}^{2M} = A_{\ell,\ell-2}^{2M} &=& \frac{w_{2M} (C_{\ell-2} + C_{\ell})}{(4\pi)^{1/2}}
\end{eqnarray}
In the quadrupolar case, the most general anisotropic two-point function is parameterized by two $\ell$-dependent
quantities ($A_{\ell-2,\ell}^{2M}$ and $A_{\ell\ell}^{2M}$), in contrast to the dipole case.
A short calculation shows that the quadrupolar anisotropy in the power spectrum is proportional to
$2 A_{\ell-2,\ell} + A_{\ell\ell}$, so that a model which satisfies $A_{\ell\ell} \approx -2 A_{\ell-2,\ell}$
has a roughly isotropic power spectrum, even though the two-point function contains a component which transforms
under rotations as a quadrupole.

Our final example is the anisotropic early universe model from \cite{ackerman/etal:2007}, in which the
initial adiabatic curvature fluctuation $\zeta({\bf k})$ is modulated in Fourier space as follows:
\begin{equation}
\zeta({\bf k}) = \left[ 1 + \sum_{M=-2}^2 w_{2M} Y_{2M}({\bf \hat k}) \right] \zeta({\bf k})_{\rm iso}
\label{eq:primordial_quadrupole}
\end{equation}
The bipolar power spectrum of this model is
\begin{equation}
A_{\ell_1\ell_2}^{2M} = \frac{i^{\ell_1-\ell_2}}{(4\pi)^{1/2}} w_{2M} 
   \int \frac{2k^2\,dk}{\pi} \Delta_{\ell_1}(k) \Delta_{\ell_2}(k) P(k)  \label{eq:primordial_quadrupole_bps}
\end{equation}
where $\Delta_\ell(k)$ is the angular CMB transfer function.

\subsection{Estimators: General Construction}

In this appendix we will construct estimators for the bipolar power spectrum and related quantities.
We will assume that the bipolar power spectrum of the CMB is a linear combination of $N$ ``template'' shapes
$A_1, A_2, \cdots A_N$.
\begin{equation}
A_{\ell_1\ell_2}^{LM} = \sum_{i=1}^N t_i (A_i)_{\ell_1\ell_2}^{LM}
\end{equation}
where the coefficients $t_i$ are to be estimated from data.  (The choice of template shapes will be discussed
shortly.)  We assume that our template shapes satisfy $A_{\ell\ell}^{00} = 0$,
i.e., the templates parameterize
deviations from isotropy, not changes in the power spectrum.

A lengthy but straightforward calculation shows that the minimum-variance estimator ${\hat t}_i$ for the template
coefficients which is unbiased (i.e., $\langle {\hat t}_i \rangle = t_i$) is given by 
\begin{equation}
{\hat t}_i[a] = F_{ij}^{-1} ({\mathcal E}_j[a] - {\mathcal N}_j)  \label{eq:ht_def}
\end{equation}
where the quantity ${\mathcal E}_j[a]$ is defined by
\begin{eqnarray}
{\mathcal E}_j[a] &=& \frac{1}{2} \sum_{\ell_1m_1\ell_2m_2LM} (A_j)_{\ell_1\ell_2}^{LM*} 
                          \sqrt{(2\ell_1+1)(2\ell_2+1)(2L+1)} \threej{\ell_1}{\ell_2}{L}{0}{0}{0} \nn \\
                   && \hspace{1cm} \times \threej{\ell_1}{\ell_2}{L}{m_1}{m_2}{M} (C^{-1}a)_{\ell_1 m_1}^* (C^{-1}a)_{\ell_2 m_2}^*
\end{eqnarray}
and the matrix $F_{ij}$ and vector ${\mathcal N}_j$ are defined by
\begin{eqnarray}
F_{ij} &=& \mbox{Cov}({\mathcal E}_i[a], {\mathcal E}_j[a])  \\
{\mathcal N}_j[a] &=& \langle {\mathcal E}_j[a] \rangle
\end{eqnarray}
where the covariance and expectation value are taken over isotropic realizations of the noisy CMB $a_{\ell m}$.

The estimator in Equation~(\ref{eq:ht_def}) can be specialized to measure different types of statistical isotropy
by making different choices of template shapes $(A_i)$.
For example, consider the dipole modulation, parameterized as in
Equation~(\ref{eq:dipole_modulation}) by a three-vector $w_i$.
We assume that multipoles $2\le\ell\le \ell_{\rm mod}$ are modulated, and multipoles $\ell > \ell_{\rm mod}$ are unmodulated.
The bipolar power spectrum is given by
\begin{equation}
A_{\ell_1\ell_2}^{LM} = \sum_{i=1}^3 w_i (A_i)_{\ell_1\ell_2}^{LM}
\end{equation}
where the template shapes $A_i$ are defined by
\be
(A_i)_{\ell_1\ell_2}^{LM} = \left\{ \barr{cl}
  (\sqrt{4\pi}/3) (C_{\ell_1} + C_{\ell_2}) Y_{1M}^*(\hat e_i) & \mbox{if $|\ell_1-\ell_2|=1$ and $\ell_i \le \ell_{\rm mod}$}  \\
  0 & \mbox{otherwise}
\earr \right.  \label{eq:dipole_template_shapes}
\end{equation}
When specialized to these three template shapes, Equation~(\ref{eq:ht_def}) gives the optimal quadratic estimator ${\hat w}_i$ for 
the three components of the modulation.  (A closely related estimator has also been constructed in 
\cite{dvorkin/peiris/hu:2008}.)
We also construct an estimator $\kappa_1$ for the total amplitude of the dipole modulation, irrespective of its direction, by
\begin{equation}
\kappa_1 = \sum_i {\hat w}_i^2
\end{equation}
We have used this estimator in Section \ref{hemispherical} to assess statistical significance of the dipole modulation, by comparing the
\iWMAP value of $\kappa_1$ to an ensemble of simulations.

The quadrupole modulation can be treated analogously.  For example, in Fig. \ref{qm} we have shown estimates of
$A_{\ell-2,\ell}^{20}$ and $A_{\ell\ell}^{20}$ in each bin, by taking $(10 N_{\rm bins})$ template shapes, corresponding
to the five components of the two bipolar power spectra in each $\ell$ bin.
As another example, if an optimal estimator (unbinned in $\ell$) for the primordial modulation in 
Equation~(\ref{eq:primordial_quadrupole}) is desired, one would take five template shapes corresponding 
to the five components of $w_{2M}$, given by Equation~(\ref{eq:primordial_quadrupole_bps}).

\subsection{Relation to likelihood formalism}

We conclude this appendix by showing how the optimal estimator is related to the maximum likelihood formalism,
for the dipole and quadrupole cases.

First consider the dipole modulation.
The likelihood function $\L[a|w]$ for the modulation $w_i$, given noisy CMB data $a_{\ell m}$, is given by:
\begin{equation}
\L[a|w] = (\mbox{const.}) \times \mbox{Det}^{-1/2}[M(w) S M(w) + N] \exp\left(-\frac{1}{2} a^\dagger [M(w) S M(w) + N]^{-1} a \right)
\end{equation}
where $M(w)$ is the operator which applies the modulation to a harmonic-space map, defined by
\begin{equation}
(M(w)a)_{\ell m} = \int d^2\n\, Y_{\ell m}^*(\n) \left( 1 + w_i \n_i \right) \left( \sum_{\ell'm'} a_{\ell'm'} Y_{\ell'm'}(\n) \right)
\end{equation}
We will show that the modulation $w$ which maximizes the likelihood $\L[w|a]$ is equal to value of the optimal quadratic estimator
$\hat w_i$ defined in the previous appendix, under two approximations that will be discussed further.

First suppose that the maximum likelihood modulation is small, so that the Taylor expansion of $(\log\L)$ to second order in $w_i$
is an accurate approximation near maximum likelihood.  The Taylor expansion is given by:
\begin{equation}
\log\L[a|w] \approx -\frac{1}{2} {\mathcal H}_{ij}[a] w_i w_j + {\mathcal G}_i[a] w_i + (\mbox{const.})  \label{eq:taylor}
\end{equation}
where we have defined:
\begin{eqnarray}
{\mathcal H}_{ij}[a] &=& -\left( \frac{\partial^2}{\partial w_i \partial w_j} \right)_{w=0} (\log\L[a|w])  \\
  &=& -a^\dagger C^{-1} M_i S M_j C^{-1} a + a^\dagger C^{-1} \{M_i,S\} C^{-1} \{M_j,S\} C^{-1} a \nn \\
   && \hspace{1cm} + \mbox{Tr}(C^{-1} M_i S M_j) - \frac{1}{2} \mbox{Tr}(C^{-1} \{M_i,S\} C^{-1} \{M_j,S\}) \\
{\mathcal G}_i[a] &=& a^\dagger C^{-1} M_i S C^{-1} a - \mbox{Tr}\left[ C^{-1} M_i S \right]
\end{eqnarray}
and $M_i = \partial M(w)/\partial w_i$.
Second, we assume that we can approximate the $a_{\ell m}$-dependent quantity ${\mathcal H}_{ij}[a]$ by its expectation value:
\begin{equation}
{\mathcal H}_{ij}[a] \approx \langle {\mathcal H}_{ij} \rangle = \frac{1}{2} \mbox{Tr}(C^{-1} \{M_i,S\} C^{-1} \{M_j,S\})
\end{equation}
Under these approximations the maximum likelihood modulation is given by
\begin{equation}
(w_i)_{\rm ML} = \langle {\mathcal H}_{ij} \rangle^{-1} {\mathcal G}_j[a]
\end{equation}
We would like to compare this expression to the optimal quadratic
estimator ${\hat w}_i$ defined in Equation~(\ref{eq:ht_def}), in the special
case where the template shapes are given by Equation~(\ref{eq:dipole_template_shapes}).  In this special case, a short calculation shows that 
$\langle {\mathcal H}_{ij} \rangle = F_{ij}$, and ${\mathcal G}_j[a] = {\mathcal E}_j[a] - {\mathcal N}_j[a]$, 
which implies $(w_i)_{\rm ML} = {\hat w}_i$.

We comment briefly on the two approximations made in this appendix,
namely the second-order Taylor approximation in Equation~(\ref{eq:taylor})
and the approximation ${\mathcal H}_{ij}[a] \approx \langle {\mathcal H}_{ij} \rangle$.  By the central limit theorem, both approximations are 
expected to become accurate in the limit where the number of CMB modes which contribute to the estimators is large.  The analyses in
Sections \ref{hemispherical} and \ref{quadrupolar} have taken wide $\ell$ bins (either the $\ell$ range $2\le \ell \le \ell_{\rm mod}$
or a series of bins with $\Delta\ell = 50$) so these approximations should be very accurate and there should be little difference in
practice between a likelihood estimator and the optimal quadratic estimator, although we have not tested this directly.
In any case, the optimal quadratic estimator has several important
advantages over a likelihood estimator, as described in Section \ref{hemispherical}.

\end{document}